%% file: clausen.tex
\documentclass{aa}             

\input epsf

\begin{document}

\def\V6{V636\,Cen}
\def\kms{\ifmmode{\rm km\thinspace s^{-1}}\else km\thinspace s$^{-1}$\fi}
\def\cm{\ifmmode{\rm cm\thinspace^{-1}}\else cm\thinspace$^{-1}$\fi}
\def\Msun{$M_{\sun}$}
\def\Rsun{$R_{\sun}$}
\def\Lsun{$L_{\sun}}
\def\feh{$[\mathrm{Fe/H}]$}
\def\meh{$[\mathrm{M/H}]$}
\def\afe{$[\mathrm{\alpha/Fe}]$}
\def\ione{\,{\sc i}}
\def\itwo{\,{\sc ii}}

\title{
Absolute dimensions of solar-type eclipsing binaries. II
\thanks{Based on observations carried out at the Str{\"o}mgren Automatic 
Telescope (SAT), the Danish 1.54m telescope, and the
1.5m telescope (62.L-0284) at ESO, La Silla, Chile.}
}
\subtitle{V636\,Centauri: \\
 A 1.05 $M_{{\small \sun}}$ primary with an active, cool, oversize 
0.85 $M_{{\small \sun}}$ secondary.
\thanks{Table~\ref{tab:v636_rv} will be available on electronic 
form at the CDS via anonymous ftp to 130.79.128.5 or via 
http://cdsweb.u-strasbg.fr/Abstract.html.}
}
\author{
J.V. Clausen     \inst{1}
\and
H. Bruntt        \inst{1,2}
\and
A. Claret        \inst{3}
\and
A. Larsen        \inst{1}
\and
J. Andersen      \inst{1,4}
\and
B. Nordstr\"om    \inst{1}
\and
A. Gim\'enez      \inst{5}
}
 
\offprints{J.V.~Clausen, \\ e-mail: jvc@astro.ku.dk}

\institute{
Niels Bohr Institute, Copenhagen University,
Juliane Maries Vej 30,
DK-2100 Copenhagen {\O}, Denmark
\and
Sydney Institute for Astronomy, School of Physics, University of Sydney,
               NSW 2006, Australia 
\and
Instituto de Astrof\'isica de Andaluc\'ia, CSIC,
Apartado 3004, E-18080 Granada, Spain
\and
Nordic Optical Telescope Scientific Association, Apartado 474, ES-38\,700
Santa Cruz de La Palma, Spain
\and
Centro de Astrobiologia (CSIC/INTA),
E-28850 Torrejon de Ardoz, Madrid, Spain 
}

\date{Received 21 April 2009 / Accepted 9 May 2009}

\titlerunning{\V6}
\authorrunning{J.V. Clausen et al.}

\abstract
{The influence of stellar activity on the fundamental properties
of stars around and below 1 \Msun\ is not well understood.
Accurate mass, radius, 
and abundance determinations from solar-type binaries exhibiting 
various levels of activity are needed for a better insight into the 
structure and evolution of these stars.}
{We aim to determine absolute dimensions and abundances for the 
solar-type detached eclipsing binary \V6, and to perform a detailed 
comparison with results from recent stellar evo\-lu\-tio\-nary models.}
{$uvby$ light curves and $uvby\beta$ standard photometry were obtained with
the Str\"omgren Automatic Telescope, radial velocity observations with the 
CORAVEL spectrometer, and high-resolution spectra with the 
FEROS spectrograph, all at ESO, La Silla.
State-of-the-art methods were applied for the photometric and spectroscopic
analyses.}
{Masses and radii that are precise to 0.5\% have been established for 
the components of \V6. 
The 0.85 \Msun\ secondary component is moderately active with starspots
and Ca\,\itwo\ H and K emission, and the 1.05 \Msun\ primary shows signs 
of activity as well, but at a much lower level.
We derive a \feh\ abundance of $-0.20 \pm 0.08$ and similar abundances for
Si, Ca, Ti, V, Cr, Co, and Ni.
Corresponding solar-scaled stellar models are unable to reproduce \V6, 
especially its secondary component, which is $\sim 10\%$ 
larger and $\sim 400$~K cooler than predicted. 
Models adopting significantly lower mixing-length parameters $l/H_p$
remove these discrepancies, seen also for other solar-type binary components.
For the observed \feh, Claret models for $l/H_p = 1.4$ (primary) and 1.0 
(secondary) reproduce the components of \V6\ at a common age of 1.35 Gyr.
The orbit is eccentric ($e = 0.135 \pm 0.001$), and apsidal motion with a 
40\% relativistic contribution has been detected. The period is 
$U = 5\,270 \pm 335$ yr, and the inferred mean central density concentration 
coefficient, log($k_2$) = $-1.61 \pm 0.05$, agrees marginally with 
model predictions.  
The measured rotational velocities, $13.0 \pm 0.2$ (primary) and 
$11.2 \pm 0.5$ (secondary) \kms, are in remarkable agreement with the 
theoretically predicted pseudo-synchronous velocities, but are about 15\% 
lower than the periastron values.
}
{\V6\ and 10 other well-studied inactive and active solar-type binaries suggest
that chromospheric activity, and its effect on envelope convection, 
is likely to cause radius and temperature discrepancies, which can be removed by
adjusting the model mixing length parameters downwards. 
Noting this, the sample may also lend support to theoretical 
2D radiation hydrodynamics studies, which predict a slight decrease of the mixing 
length parameter with increasing temperature/mass for $\it inactive$ main sequence stars.
More binaries are, however, needed for a description/calibration in terms
of physical parameters and level of activity.
}
\keywords{
Stars: evolution --
Stars: fundamental parameters --
Stars: activity --
Stars: binaries: eclipsing --
Techniques: spectroscopic}

\maketitle

\section{Introduction}
\label{sec:intro}

As pointed out by e.g.
Hoxie (\cite{h73}),     
Popper (\cite{dmp97}), 
Clausen et al. (\cite{granada99}),
Torres \& Ribas (\cite{tr02}),
Ribas (\cite{ir03}),
Dawson \& De Robertis (\cite{ddr04}), and
L\'opez-Morales \& Ribas (\cite{lmir05}),
current stellar models, scaled to the Sun, are unable to reproduce the 
measured temperatures and radii for many binary components less massive 
the Sun. Such models tend to predict too high temperatures, and radii, which are
up to 10\% too small, compared to a measured accuracy of about 1\% for several systems.
The model luminosities seem, however, to agree fairly well with observations.
Chromospheric activity has been suggested as a likely cause, and e.g.
Gabriel (\cite{g69}),         
Cox et al. (\cite{csh81}),
Clausen et al. (\cite{granada99}), and
Torres et al.(\cite{wt06}) have demonstrated that the model fits 
can be improved by using a reduced mixing-length parameter when 
calculating envelope convection. The latter authors also show that there is 
a link between activity and increased radii.
Morales et al. (\cite{mrj08}) find strong radius and 
temperature dependencies on stellar activity for both single and
binary low-mass stars, and L\'opez-Morales (\cite{lm07}) finds
correlation between activity level (X-ray emission level) and 
radius for faster rotating binary components, but not for slowly 
rotating single stars.
On the theoretical side, Chabrier et al. (\cite{cgb07}) have,
via a phenomenological approach, examined the consequences on the 
evolution of low-mass stars and brown dwarf eclipsing binaries of 
a) inhibiting convection due to rotation and/or internal magnetic field, 
and b) the presence of surface magnetic spots.

The current situation is summarized by Ribas et al. (\cite{rmjbcg08}).
Clearly, accurate data for additional binaries are needed, as well as
more models, which include first of all dynamo magnetic fields and the
evolution of stellar rotation and activity (e.g. D'Antona et al.
\cite{da00}).
We are presently studying several new systems with solar-type components, 
either constant or exhibiting various levels of chromospheric activity
(e.g. Clausen et al. \cite{jvcetal01}), and in this paper, we present 
results from a complete analysis of the F8/G0~V\footnote{Houk (\cite{houk78}); 
spectral type of primary component.} system \object{\V6} $=$ HD124784 $=$ 
HIP69781.

\section{\V6}
\label{sec:v636cen}

\V6\ was discovered by Hoffmeister (\cite{h58}) to be a $4\fd3$ period 
eclipsing binary, and Popper (\cite{dmp66}) reported sharp but single lines. 
Today \V6 is known to be double-lined, but the lines of the secondary 
component are much fainter than those of the primary. 
\V6 is well detached with two quite different components in an eccentric 
orbit ($e = 0.135$). Apsidal motion with a period of $U = 5\,270 \pm 335$ yr 
has been detected (Clausen et al. \cite{jvcetal08}, hereafter CVG08). 
Signs of chromospheric activity (spots) are clearly present in the light 
curves, but the level of the light variations is modest (CVG08). 
Furthermore, we see emission from both 
components in the Ca\,\itwo\ H and K lines, whereas $\mathrm{H_{\alpha}}$ 
emission is not noticed; see Sect.~\ref{sec:phel}.
Chromospheric emission in the Ca\,\itwo\ H and K lines was also observed 
by Henry et al. (\cite{henry96}), and X-ray emission was detected 
in the $ROSAT$ All-Sky Survey (Voges et al. \cite{rosat99}).

Our study is based on the first modern light curves and spectroscopic 
observations for \V6; preliminary dimensions based on part of the data 
were derived by Larsen (\cite{al98}).
For the spectroscopic and photometric analyses of \V6 presented below, we 
have adopted the linear ephemeris by CVG08 (Eq. 4). 
Throughout the paper, the component eclipsed at the deeper eclipse at phase 0.0 
is referred to as the primary $(p)$, and the other as the secondary $(s)$ 
component.

\section{Photometric elements}
\label{sec:phel}

The $uvby$ light curves of \V6\ (CVG08) contain 853 observations in each band 
and were observed on 76 nights during six periods between March 1985 and April 
1991. The number of observations per season are 8 (1985), 165 (1987),
261 (1988), 233 (1989), 17 (1990), and 169 (1991). Further eclipse observations,
but none outside eclipses, were done on ten nights between January 2002 
and July 2007, primarily to determine the apsidal motion period of the eccentric
orbit. 
Primary eclipse is much deeper than secondary eclipse, which is total and occurs
near phase 0.52.
  
The average observational accuracy per data point is about 5 mmag ($vby$) and
7 mmag ($u$), and the magnitude differences between the two comparison stars 
were found to be constant at that level. 
Additional scatter due to chromospheric activity (spots) is, however, clearly 
seen in the light curves, with total amplitudes increasing from about 0.02 
mag in $y$ to about 0.04 mag in $u$, and changing somewhat from year to year. 
We refer to Fig.~6 and Table~9 in GVG08 for further details.

\begin{figure}
\epsfxsize=90mm
\epsfbox{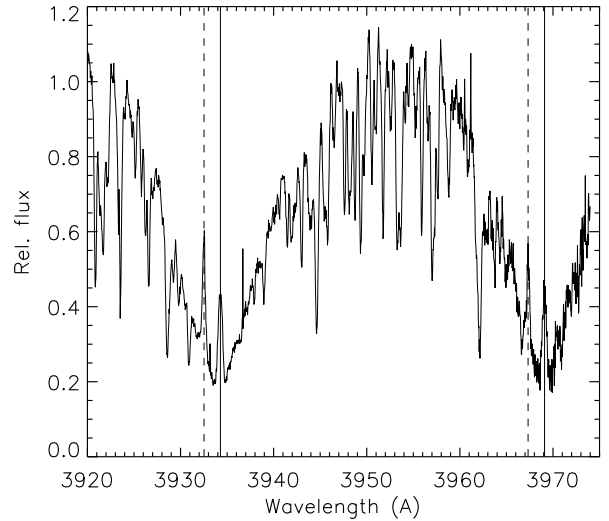}
\caption[]{\label{fig:v636cen_ca}
Ca\,\itwo\ H and K emission in \V6.
The vertical lines show the positions for the primary (full)
and secondary (dashed) components.
The FEROS spectrum of \V6 was observed at JD2451211.85 (orbital phase 0.63).}
\end{figure}

\input 8788_ebop_vh.tex

As mentioned in Sect.~\ref{sec:intro}, Ca\,\itwo\ H and K emission 
is present from both components. This is seen in two orders of 
two FEROS spectra from February 1999 (Table~\ref{tab:feros}), 
and (dark) spots may therefore be present at the surface of both stars. 
Fig.~\ref{fig:v636cen_ca} illustrates that the emission features of the
two components appear to have about the same strength, but since the primary
dominates the total light of \V6, the real emission from the secondary
is by far the strongest.

Between 1987 and 1991, the depths of the total secondary eclipses were nearly 
identical, whereas those of the primary eclipses differed significantly. 
During this period, the activity must therefore have been dominated by 
contributions from the secondary component. As mentioned below, this is 
supported by frequency analyses of the out-of-eclipse data. 
On the other hand, the depths of both eclipses varied from 2002 to 2006.
This is not due to the slow apsidal motion detected for \V6, and must
be caused by spots at both stars during at least some of those years.
The Rossby\footnote{Defined as the ratio of the rotation 
period to the convective turnover time.} numbers for the primary and 
secondary components are approximately 0.44 and 0.10, respectively. This 
places the secondary in the range where other stars tend to show spot 
activity and the primary close to onset of spottedness (Hall \cite{hall94}).

\input 1987_wd_vh.tex
\input 1988_wd_vh.tex

Out-of-eclipse phases were covered on several nights during 1987, 1988,
1989, and 1991. Frequency analyses of the corresponding data, done
separately for each year and band, reveal the existence of one significant 
period per year. The results from the four bands agree
well, and the average periods and estimated uncertainties are 
$4\fd05 \pm 0\fd06$ (1987), 
$4\fd16 \pm 0\fd01$ (1988),
$4\fd13 \pm 0\fd01$ (1989), 
and $4\fd06 \pm 0\fd12$ (1991). 
They are all shorter than the orbital period $4\fd28$, but transformed
to equatorial surface velocities they agree quite well with the observed
rotational velocities of both components (Sect.~\ref{sec:specorb}). 

Based on this information on eclipse depths and periodicity of the 
out-of-eclipse variability, we conclude that the 1987-1991 surface activities 
can be modeled adequately by including a few dominating spots (or spot groups) 
on the surface of the secondary component. 

We have used the Wilson-Devinney (WD) model and code for the analyses. 
As described below, some initial analyses, which ignore surface activity, 
were based on the JKTEBOP program. 
We refer to Clausen et al. (\cite{avw08}, hereafter CTB08) for references 
and details on the binary models and codes, and on the general approach 
applied. 
In tables and text, we use the following symbols:
$i$ orbital inclination;
$r$ relative radius;
$k = r_s/r_p$;
$\Omega$ surface potential;
$u$ linear limb darkening coefficient;
$y$ gravity darkening coefficient;
$J$ central surface brightness;
$L$ luminosity;
$T_{\rm eff}$ effective temperature.

\begin{figure*}
\epsfxsize=180mm
\epsfbox{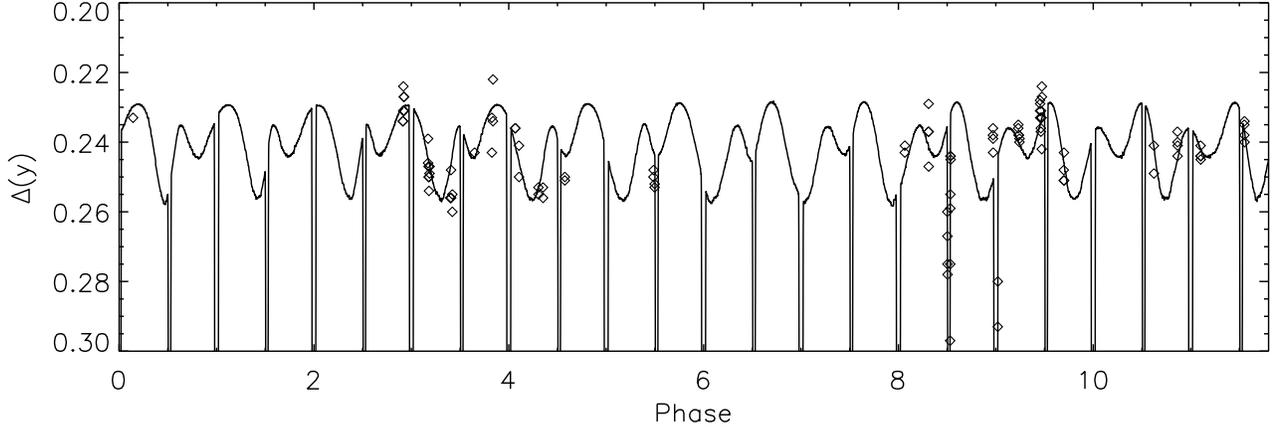}
\caption[]{\label{fig:1987y}
Fit of the theoretical light curve of the WD solution from 
Table~\ref{tab:1987_wd_vh} to the 1987 $y$ observations.
}
\end{figure*}

\begin{figure*}
\epsfxsize=180mm
\epsfbox{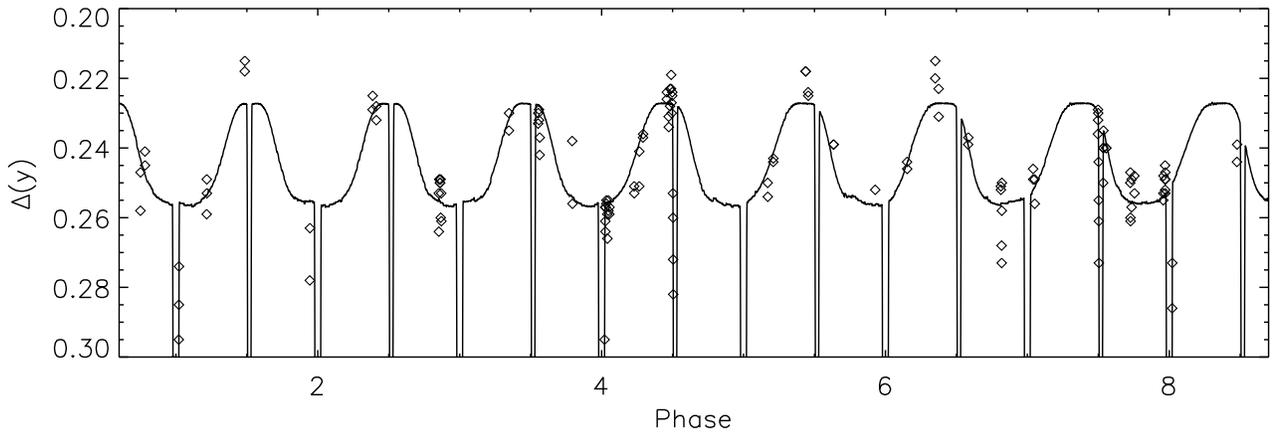}
\caption[]{\label{fig:1988y}
Fit of the theoretical light curve of the WD solution from 
Table~\ref{tab:1988_wd_vh} to the 1988 $y$ observations.
}
\end{figure*}

A linear limb darkening law has been assumed throughout with coefficients 
adopted from Van Hamme (\cite{vh93}) and/or Claret (\cite{c00}). 
They differ by 0.02--0.10 only, leading to identical photometric elements
within errors.
Gravity darkening coefficients/exponents corresponding to convective 
atmospheres were applied, and bolometric reflection albedo coefficients of 0.5
were chosen in the WD analyses, again due to convection. For JKTEBOP, the 
simple bolometric reflection model was used.
The mass ratio between the components was kept at the spectroscopic value
($M_s/M_p=0.812\pm0.002$). In order to model the spots properly, the 
(very nearly spherical) components were assumed to rotate at the periods 
determined from the frequency analyses mentioned above. 
As initial analyses showed that the orbital eccentricity is not as well
constrained from the light curve analyses as from the spectroscopic orbit,
we have adopted the spectroscopic result throughout; 
see Table~\ref{tab:v636_orb}.
The longitude of periastron $\omega$ is then defined through $e \cos \omega$ 
from the well determined phase of central secondary eclipse, which did not change
significantly relative to primary eclipse during the light curve observations 
from 1987 to 1991.

As the activity level of \V6 is modest, it is appropriate to first analyse 
the light curves without taking spots into account, and to adopt a simple 
but adequate binary model/code like JKTEBOP, which allows easy determination 
of realistic errors of the photometric elements from Monte Carlo simulations. 
Since the average out-of-eclipse level of the 1991 observations is 
significantly off, and since there are very few 1985 and 1990 observations, 
these data sets were not included in the analyses. 
Results from the combined 1987 and 1988 observations, which both cover all 
phases well, are given in Table~\ref{tab:8788_ebop_vh}. Including also
the 1989 observations increases the scatter significantly but leads to
identical solutions within errors.
As seen, the results from the four bands agree remarkably well, and relative 
radii with a realistic precision of about 1\% are obtained. 
Changing the adopted orbital eccentricity by +0.005, i.e. four times the error 
of the spectroscopic result, decreases the relative radii of both components 
by only 0.5\%. 
Adjusting $e$ and $\omega$ leads to spurious results, due to
asymmetries in the observed light curves caused by spots.
As expected, photometric elements from similar WD analyses, not accounting for 
spots, agree very well with those from JKTEBOP.

Results from WD analyses of the 1987 and 1988 observations are
presented in Tables~\ref{tab:1987_wd_vh} and \ref{tab:1988_wd_vh},
and out-of-eclipse comparisons in the $y$ band are shown in 
Figs.~\ref{fig:1987y} and \ref{fig:1988y}.
The activity is modeled by including two rather big spots at the surface of
the secondary component. Most probably, this simplified approach does 
not represent the real surface structure in detail, but it reproduces the light 
variations quite closely, and somewhat better than if only one spot is
included. Based on extensive tests, the spots were placed at the equator of 
the secondary component, and a ratio of 0.9 between the temperatures of 
spotted and unspotted areas was selected. Longitudes ($LO$) and 
angular radii ($AR$) 
of the spots were adjusted together with orbital and stellar parameters.

As seen, the eight independent WD solutions yield nearly identical relative
radii of the components, which are systematically about 1.5\% smaller
than the JKTEBOP results (Table~\ref{tab:8788_ebop_vh}). 
The orbital inclinations of the 1987 solutions are on average about
$0\fdg25$ higher than those from the 1988 solutions, but tests show that
the realistic error of $i$ is at least 3-4 times larger than the formal
errors listed.
The luminosity ratios given in Tables~\ref{tab:1987_wd_vh} and
\ref{tab:1988_wd_vh}, which for the sake of comparison do not include 
the spots, agree well within realistic errors; we will return to
this point below.
For each year, the spot parameters from the four bands differ slightly, but 
not more than can be expected from the simple approach we have adopted and the
limited number of observations available. 
The two spots applied cover about 10\% of the surface of the secondary
component.
Finally, the rms errors ($\sigma$) of the fits to the observations are close 
to the observational errors per data point.

Similar WD analyses of the 1989 observations were not successful.
The ascending branch of secondary eclipse is not covered, and the
level of activity seems to have changed during the last half of the
observing period. Secondary eclipse was not covered in 1991,
and very few 1985 and 1990 observations exist, so these data sets were
not analysed.

\input v636_phel.tex

As a further test of the simple spot model adopted, it is of interest
to compare the 1987 and 1988 light curves cleaned for spot effects.
This is easily done by calculating, year by year and band per band, 
theoretical light curves with and without spots and then subtract the 
difference between them from the observations. 
First, the cleaned 1987 and 1988 light curves agree well, and second,
JKTEBOP analyses of the combined cleaned data yield stellar radii, which are
now in perfect agreement with those from the WD solutions. The orbital
inclination is close to the mean of the 1987 and 1988 WD results.

The adopted photometric elements are presented in Table~\ref{tab:v636_phel}.
The main result of the photometric analyses is that we have, in spite of the
surface activity, been able to determine very accurate relative radii for
the components of \V6. The system is well detached and the stars are quite
different, especially with respect to temperature and luminosity. 
As seen, the orbital inclination is high, and consequently 
secondary eclipse is total.
The luminosity ratios listed in Table~\ref{tab:v636_phel} are based on the
mean stellar and orbital parameters and include the average spot level seen in 
the 1987 and 1988 observations; errors reflect the difference between the
two years. This approach was adopted in order to allow calculation of realistic
$uvby$ indices for the components; see Table~\ref{tab:v636_absdim}.
Due to the spots on the surface of the secondary component, the ratio of the
light contributions outside eclipses of course change significantly through 
the orbit. However, the $b-y$, $m_1$, and $c_1$ indices of the components 
remain nearly unaffected.

\section{Spectroscopic orbit}
\label{sec:specorb}

\begin{figure}
\epsfxsize=095mm
\epsfbox{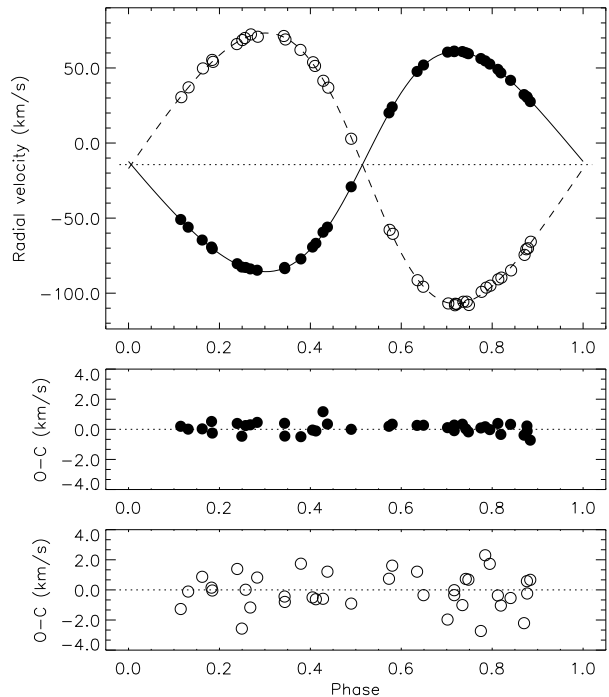}
\caption[]{\label{fig:v636_orb}
Spectroscopic orbital solution for \V6\ (solid line: primary;
dashed line: secondary) and radial velocities (filled circles:
primary; open circles: secondary).
The dotted line (upper panel) represents the center-of-mass
velocity of the system.
Phase 0.0 corresponds to central primary eclipse.
}
\end{figure}

\input v636_orbit.tex

The radial velocity measurements of \V6 were obtained with the photoelectric 
cross-correlation spectrometer CORAVEL (Baranne et al. \cite{b79}; 
Mayor \cite{m85}) operated at the Danish 1.5-m telescope at ESO, La Silla.
In total, 38 and 39 measurement were obtained for the primary and secondary 
components, respectively, on 30 nights between March 1986 and June 1989. 
The internal velocity errors range between 0.39 and 0.65 \kms\
(median 0.44 \kms) for the primary component 
and between 0.71 and 1.81 \kms\ (median 0.88 \kms) for the 
fainter secondary component.
The observations, which cover the out-of-eclipse phases well, 
are listed in Table~\ref{tab:v636_rv}.

The spectroscopic orbits were calculated using the method of 
Lehman-Filh\'es implemented in the SBOP\footnote{Spectroscopic 
Binary Orbit Program, \\ {\tt http://mintaka.sdsu.edu/faculty/etzel/}}
program (Etzel \cite{sbop}), which is a modified and expanded version of 
an earlier code by Wolfe, Horak \& Storer (\cite{wolfe67}).
Following several tests, we have decided to weight the individual radial 
velocities according to the inverse square of their internal errors.
The SB1 and SB2 solutions for \V6 are given in Table~\ref{tab:v636_orb}, and the 
observations and computed orbit are shown in Fig.~\ref{fig:v636_orb}.
As seen, the SB2 solution, which we adopt, agrees well with the individual 
SB1 solutions. 
The semiamplitudes yield minimum masses accurate to about 0.4\% for both 
components, 
and the eccentricity and periastron longitude of the orbit are well constrained.
Also, the standard deviations of a single velocity agree well with the internal
median velocity errors.

The measured $v \sin i$ values are $13.0 \pm 0.2$~\kms\
and $11.2 \pm 0.5$~\kms, respectively, with the velocity of the secondary 
being more uncertain due to its much weaker lines.
These rotational velocities are supported by the frequency analyses of
the out-of-eclipse photometry described in Sect.\ref{sec:phel}.
As discussed in Sect.~\ref{sec:absdim}, the rotation of the components
corresponds closely to pseudo-synchronization.

\section{Abundances}
\label{sec:abund}

For abundance determinations, we have obtained two high-resolution
spectra of \V6\ with the FEROS fiber echelle spectrograph at ESO,
La Silla in February 1999; see Table~\ref{tab:feros}. 
Details on the spectrograph and the reduction of the spectra are given by CTB08.
The late-type components of \V6 have rich absorption
line spectra, and unlike many known solar-type systems
(Table~\ref{tab:systems}), the components
of \V6 are relatively slow rotators. This is a clear advantage
for the abundances studies. The ideal case would be to have
enough spectra for disentangling\footnote{About 10 spectra with
S/N $\sim$ 100, well distributed in phase, would be required}, 
since mutual blending is avoided and the reconstructed component 
spectra 
get a higher signal-to-noise ratio than each of the observed spectra.
However, as will be demonstrated below, the wide spectral 
coverage of FEROS provides a large number of unblended lines 
from both components which are usable for the analyses of the combined spectra. 
Furthermore, the two spectra have been observed at 
different phases giving opposite line shifts. 
Since the primary component is 3--5 times more luminous than the secondary, 
depending on wavelength, its lines are only diluted by a factor of about 1.3, 
meaning that an accuracy close to that for a similar single star with spectra 
of comparable quality can be reached. For the secondary, the situation is on 
the other hand worse; its dilution factor is about 4.7.
A lower accuracy must therefore be expected for the secondary
due to higher errors of the intrinsic equivalent widths.
In general, also intrinsically stronger lines, for which the equivalent widths 
are less sensitive to abundance and more sensitive to e.g. microturbulence and 
pressure broadening, have to be used for the secondary.
This is in principle avoided in disentangling, but the signal-to-noise ratio
of the reconstructed spectrum for the secondary would on the other hand
still be much lower than for the primary.

The basic approach followed in the abundance analyses is described
by CTB08. The versatile VWA tool, now extended to analyses of double-lined
spectra, was used; we refer to Bruntt et al. (\cite{bruntt04},
\cite{bruntt08}) for an updated description of VWA. It uses the SYNTH 
software (Valenti \& Piskunov \cite{vp96}) to generate the synthetic spectra.
Atmosphere models were interpolated from the grid of modified ATLAS9 
models by Heiter et al. (\cite{heiter02}). 
Line information was taken from the Vienna Atomic Line Database (VALD;
Kupka et al. \cite{kupka99}), but in order to derive abundances relative
to the Sun, log($gf$) values have been adjusted in such a way that each 
measured line in the Wallace et al. (\cite{whl98}) solar atlas reproduces 
the atmospheric abundances by Grevesse \& Sauval (\cite{gs98}). 
Analyses of a FEROS sky spectrum reproduce these adjustments closely.

\input feros.tex

The abundance results derived from all useful lines in both spectra 
are presented in Table~\ref{tab:v636cen_abund}. 
Except for a few ions, we have only included lines with measured equivalent
widths above 10 m{\AA} and below 80 m{\AA} (primary) and 50 m{\AA} (secondary).
Comparing the results from the individual spectra, we find no significant 
differences. The effective temperatures, surface gravities 
and rotational velocities listed in Table~\ref{tab:v636_absdim} were adopted.
Microturbulence velocities were tuned until Fe \ione\ abundances were 
independent of line equivalent widths, and the resulting values are 1.15 
(primary) and 1.35 (secondary) \kms. The calibration by Edvardsson et
al. (\cite{be93}) predicts $1.25 \pm 0.31$ \kms\ for the primary component, 
whereas the secondary lies outside its effective temperature range.
For the adopted effective temperatures we see no dependency of the
abundance on excitation potential, which, however, occurs if they are changed
by more than $\pm 100-150$~K.

As seen, a robust \feh\ is obtained for the primary, with identical
results from Fe\,\ione\ and Fe\,\itwo\ lines. Within errors, the
less precise result for the secondary from Fe\,\ione\ lines agrees;
unfortunately no suitable, unblended Fe\,\itwo\ lines are available.
Changing the primary model temperatures by $\pm 100$~K modifies \feh\ 
from the Fe\,\ione\ lines by $\pm0.05$ dex, and a similar but
opposite effect is seen for Fe\,\itwo\ lines.
If 0.3 \kms\ higher microturbulence velocities are adopted,
\feh\ decreases by about 0.05 dex for both neutral and ionized lines.
Finally, changing the adopted luminosity ratio between the
components (Table~\ref{tab:v636_phel}) by $+1$~$\sigma$ changes \feh\
by +0.015 dex for the primary and -0.04 dex for the secondary.
Taking these contributions to the uncertainties into account, 
and giving higher weight to
the primary, we adopt \feh\,$=-0.20\pm0.08$ for \V6.
Except for Cr\,\itwo\, we also find relative abundances close to -0.20 dex 
for the other ions listed in Table~\ref{tab:v636cen_abund}, including the
$\alpha$-elements. The reason for the Cr\,\itwo\ discrepancy is not clear, 
but a similar result was recently found for the G0~V system WZ\,Oph (CTB08). 

\input v636cen_abund.tex

We have also done abundance analyses based on the recent grid of MARCS
model atmospheres (Gustafsson et al. \cite{marcs08}), which adopt
the solar composition by Grevesse et al. (\cite{gas07}). 
Adjusting the log($gf$) accordingly, as described above, we obtain
abundances for \V6\ which agree with those listed in 
Table~\ref{tab:v636cen_abund} within about $\pm 0.05$ dex. 

As a supplement to the spectroscopic abundance analyses, we have
derived metal abundances from the de-reddened $uvby$ indices for the
individual components (see Table~\ref{tab:v636_absdim}) and 
the calibration by Holmberg et al. (\cite{holmberg07}).
For the primary component, the result, \feh\,$=-0.23\pm0.11$,
agrees perfectly well with the spectroscopic determination, whereas an
unrealistically low value, \feh\,$=-0.77\pm0.17$, is obtained
for the secondary. Their previous calibration for late type
stars (Nordstr\"om et al. \cite{gencph04}) gives \feh\,$=-0.86\pm0.20$.
The quoted \feh\ errors include the uncertainties of the photometric indices 
and the published spread of the calibrations, and the reason for the 
discrepancy between the primary and secondary results is not clear. It may, 
however, very likely be due to $m_1$ deficiency for the secondary related to
activity, as in fact seen for active binaries (Gim\'enez et al. \cite{agc91}), 
rather than a strong metal underabundance. 
Adopting the definitions by Gim\'enez et al., we find $\delta m_1 \sim 0.11$ 
and $\delta c_1 \sim 0.00$ for the secondary. If we modify $m_1$ 
accordingly and assume that $b-y$ is unaffected, the two calibrations
mentioned above lead to \feh\,$=-0.34\pm0.18$ and \feh\,$=-0.36\pm0.21$,
respectively, i.e. reasonable agreement with the result for the primary is 
reached.

\section{Absolute dimensions}
\label{sec:absdim}

\input v636_absdim.tex

Absolute dimensions for the components of \V6 are calculated
from the elements given in Tables~\ref{tab:v636_phel} and
\ref{tab:v636_orb}. As seen in Table~\ref{tab:v636_absdim},
both masses and radii have been obtained to an accuracy of about 0.5\%.

Individual standard $uvby$ indices are included in Table~\ref{tab:v636_absdim},
as calculated from the combined indices of \V6 outside eclipses (CVG08)
and the luminosity ratios (Table~\ref{tab:v636_phel}).
The calibration by Olsen~\cite{eho88}, the $uvby$ indices of the 
primary component and $\beta$ observations obtained during the total secondary 
eclipse ($\beta = 2.601 \pm 0.008$) then yield a very small interstellar 
reddening of $E(b-y) = 0.005 \pm 0.008$.

The adopted effective temperatures (5900~K, 5000~K) were calculated from the 
calibration by Holmberg et al. (\cite{holmberg07}), assuming \feh\,$=-0.20$
(Sect.~\ref{sec:abund}). The uncertainties include those of the $uvby$ indices,
$E(b-y)$, \feh\, and the calibration itself.
Temperatures based on the calibrations by Alonso et al. (\cite{alonso96})
and Ram\'irez \& Mel\'endez (\cite{rm05}) agree within errors,
but are, however, systematically about 100~K lower for the primary component.

From the orbital and stellar parameters, the observed rotational velocities,
and the apsidal motion period determined by CVG08, we derive a mean central 
density coefficient of log($k_2$) = $-1.61 \pm 0.05$. The observed rate 
of change of the longitude of periastron
$\dot\omega = 0.00080 \pm 0.00005$ \degr/c was corrected for a 40\% 
relativistic contribution of $\dot\omega = 0.00032$ \degr/c, as calculated 
from general relativity (Gim\'enez, \cite{agc85}). The higher order $k_3$ 
and $k_4$ terms are negligible.
From theoretical coefficients, calculated for the models described in 
Sect.~\ref{sec:models_lhp} along the lines described by Claret (\cite{claret95,claret97}),
we derive a theoretical log($k_2$) of $-1.74$.

As seen in Table~\ref{tab:v636_absdim}, the measured rotational velocities are 
in remarkable agreement with the theoretically predicted pseudo-synchronous velocities 
(Hut \cite{hut81}, Eq. 42), but are about 15\% lower than the periastron values.
The turbulent dissipation and radiative damping formalism of Zahn 
(\cite{zahn77,zahn89}) predicts
synchronization time scales of 0.1 Gyr for both components
and a time scale for circularization of 7.5 Gyr,
whereas the hydrodynamical 
mechanism by Tassoul \& Tassoul (\cite{tt97}, and references therein)
yields much lower values of 0.3 Myr (primary synchronization), 0.4 Myr 
(secondary synchronization), and 0.04 Gyr (circularization). 
We refer to Claret \& Cunha (\cite{ac97}) and Claret et al. (\cite{ac95}) 
for details on the calculations.
From model comparisons (Sect.~\ref{sec:models_lhp}), we derive an age for
\V6\ of about 1.35 Gyr, meaning that both theories are in
agreement with the observed pseudo-synchronization.
On the other hand, the mechanism by Tassoul \& Tassoul
predicts a circular orbit, which is certainly not the case. 

The distance to \V6\ was calculated from the "classical" relation
(see e.g. CTB08), adopting the solar values and bolometric corrections 
given in Table~\ref{tab:v636_absdim} and $A_V/E(b-y) = 4.27$,
and accounting for all error sources. 
As seen, nearly identical results are obtained from the two components.
The mean distance, 71.8 pc, which has been established to about 
4\%, is in perfect agreement with the result of the 
new Hipparcos reduction by van Leeuwen (\cite{vl07}), $72.2 \pm 4.7$ pc, but 
is marginally larger than the original Hipparcos result $65.1 \pm 4.7$ pc
(ESA \cite{hip97}). 
We note that \V6\ belongs to the group of eclipsing binaries within 125 pc, 
discussed by Popper (\cite{dmp98}), which could be useful for improving the 
radiative flux scale. 

\section{Discussion}
\label{sec:dis}

In the following, we first compare the absolute dimensions obtained 
for \V6\ with properties of recent theoretical stellar evolutionary models.
We then compare \V6\ to other well-studied eclipsing binaries with at least
one component in the 0.8--1.1 $M_{\sun}$ mass interval.

\subsection{Comparison with solar-scaled models}
\label{sec:models_sun}

\begin{figure}
\epsfxsize=90mm
\epsfbox{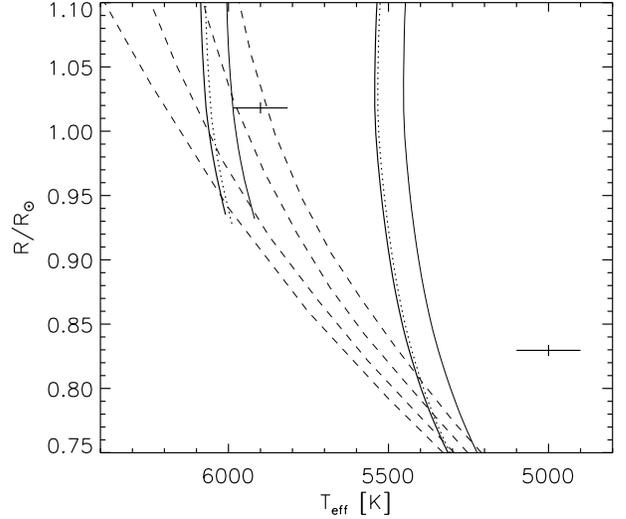}
\caption[]{\label{fig:v636cen_tr_y2}
\V6\ compared to $Y^2$ models 
calculated for the measured abundance \feh\,$=-0.20$.
Tracks for the component masses (full drawn, thick) and isochrones for 0.5,
2.0, 4.0, and 6.0 Gyr (dashed) are shown.
The uncertainty in the location of the tracks coming from
the mass errors are indicated (dotted lines).
To illustrate the effect of the abundance uncertainty, tracks for
\feh\,$=-0.12$ 
are included (full drawn, thin).
}
\end{figure}

\begin{figure}
\epsfxsize=90mm
\epsfbox{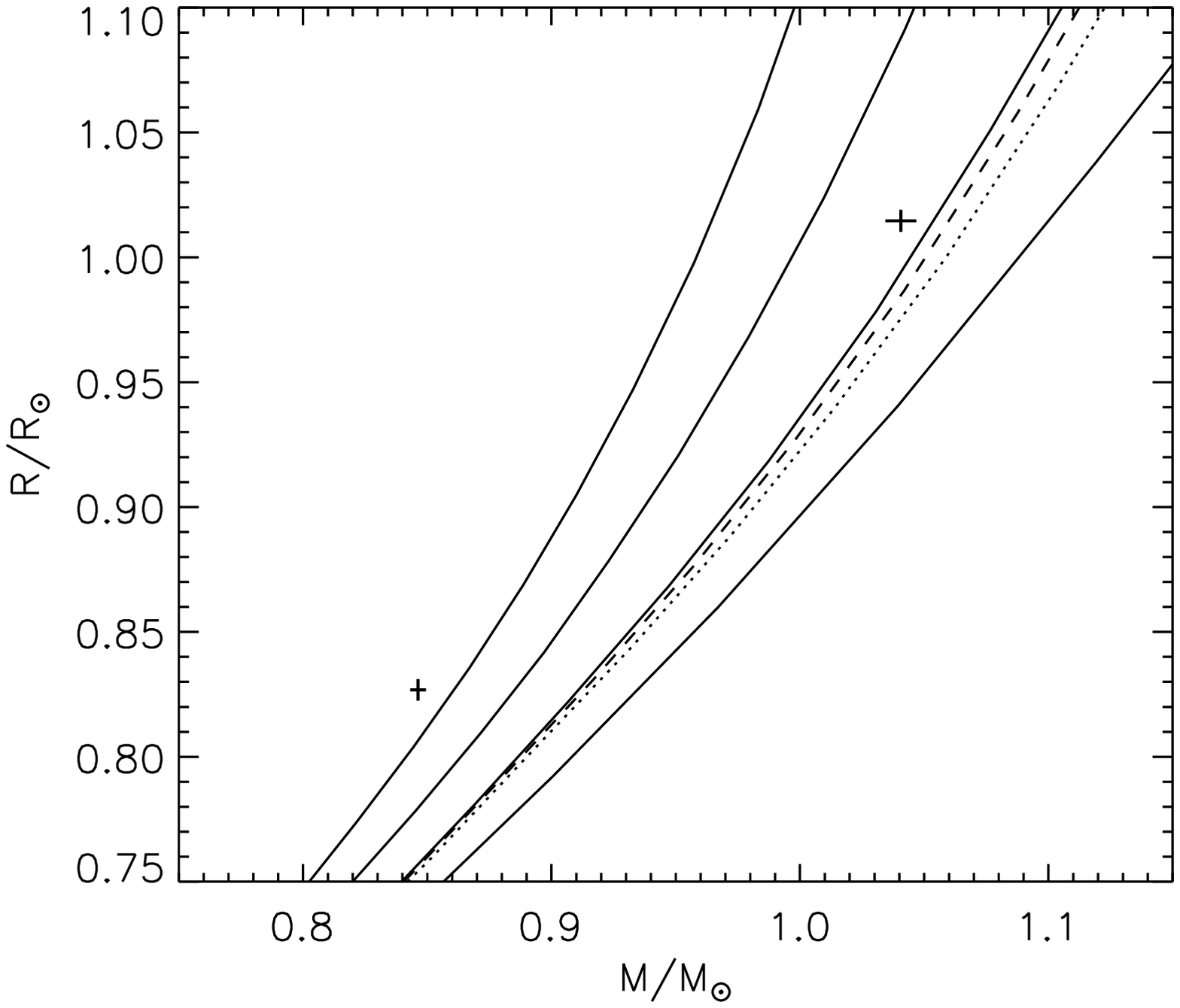}
\caption[]{\label{fig:v636cen_mr_y2}
\V6\ compared to $Y^2$ models 
calculated for \feh\,$=-0.20$. 
Iso\-chro\-nes for 0.5, 2.0, 4.0, and 6.0 Gyr are shown.
For comparison, 2 Gyr isochrones for \feh\,$=-0.12$ (dashed)
and \feh\,$=0.00$ (dotted) are included.
}
\end{figure}

\begin{figure}
\epsfxsize=90mm
\epsfbox{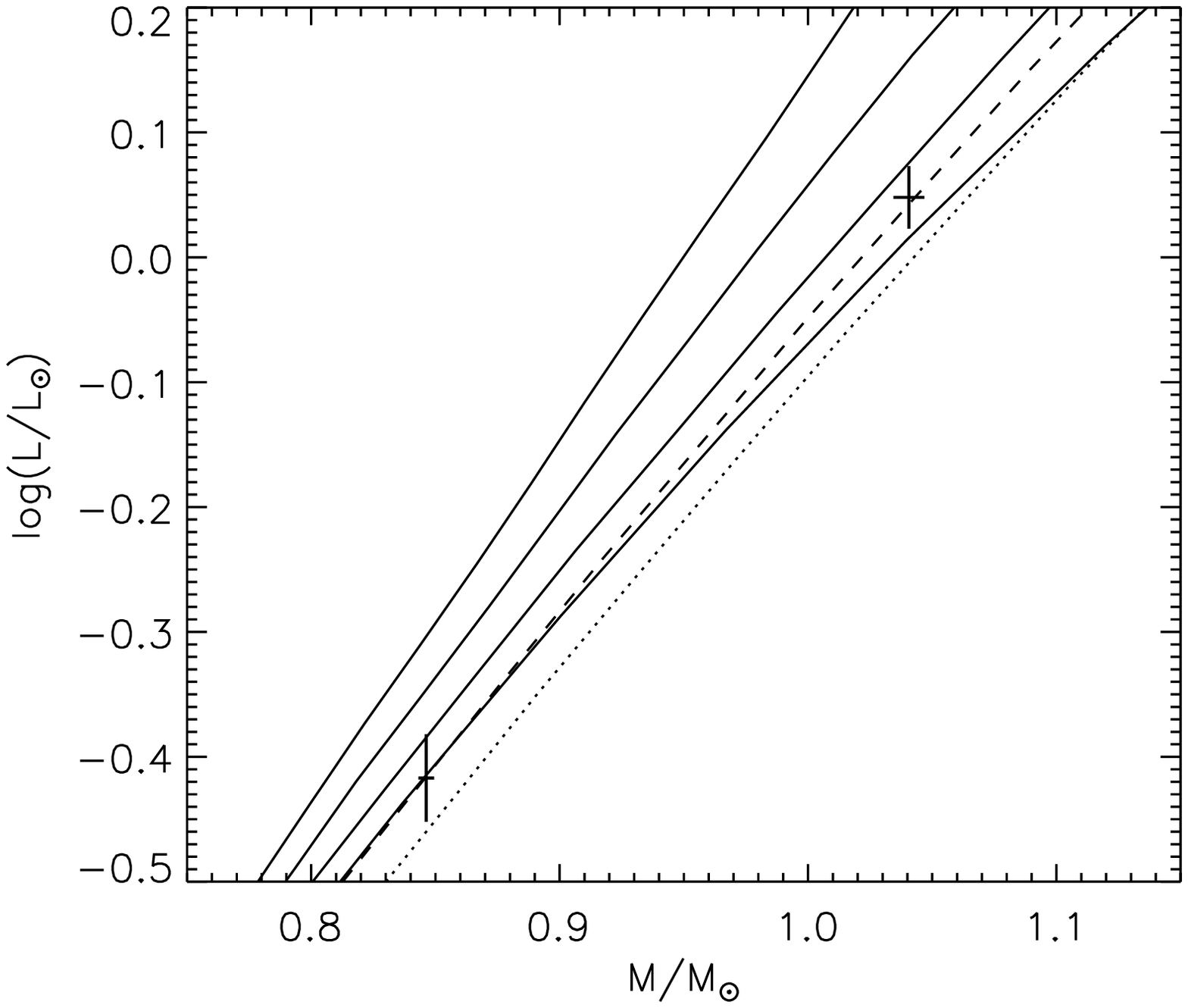}
\caption[]{\label{fig:v636cen_ml_y2}
\V6\ compared to $Y^2$ models 
calculated for \feh\,$=-0.20$.
Iso\-chro\-nes for 0.5, 2.0, 4.0, and 6.0 Gyr are shown.
For comparison, 2 Gyr isochrones for \feh\,$=-0.12$ (dashed)
and \feh\,$=0.00$ (dotted) are included.
}
\end{figure}

Figs.~\ref{fig:v636cen_tr_y2}, \ref{fig:v636cen_mr_y2}, 
and \ref{fig:v636cen_ml_y2} illustrate the results from comparisons with the 
Yonsei-Yale ($Y^2$) evolutionary tracks and isochrones by Demarque et al.
(\cite{yale04})\footnote{{\scriptsize\tt http://www.astro.yale.edu/demarque/yystar.html}}.
The mixing-length parameter in convective envelopes is calibrated using
the Sun, and is held fixed at $l/H_p = 1.7432$.
The enrichment law $Y = 0.23 + 2Z$ is adopted, together with the solar 
mixture by Grevesse et al. (\cite{gns96}), leading to
($X$,$Y$,$Z$)$_{\sun}$ = (0.71564,0.26624,0.01812).
Only models for \afe\,$=0.0$ have been considered.
We refer to CTB08 for a brief description of other aspects of their 
up-to-date input physics.

As seen from Fig.~\ref{fig:v636cen_tr_y2}, models for the observed
masses and abundance, \feh\,$=-0.20$, equivalent to
($X$,$Y$,$Z$) = (0.73475,0.25350,0.01175),
are significantly hotter than observed, 
especially for the secondary component. The uncertainty of 
\feh\ is $\pm 0.08$ dex, and models for \feh\,$=-0.12$,
equivalent to ($X$,$Y$,$Z$) = (0.72809,0.25790,0.01397),
marginally fit 
the primary but not the secondary component. In order to fit the primary 
perfectly, $Y^2$ models with a heavy element content close to solar, 
or a higher He abundance, are required. Such models would, however, still be 
unable to reproduce the secondary. 
The same picture is seen for e.g. Victoria-Regina (VandenBerg et al.,
\cite{vr06})\footnote{{\scriptsize\tt http://www1.cadc-ccda.hia-iha.nrc-cnrc.gc.ca/cvo/
community/VictoriaReginaModels/}}
and BaSTI (Pietrinferni et al., 
\cite{basti04})\footnote{{\scriptsize\tt http://www.te.astro.it/BASTI/index.php}}
models, which differ slightly from $Y^2$, e.g. with respect to input physics
and He enrichment law. We refer to CTB08 for a brief comparison of the three model grids.

The most direct comparison, based on the {\it scale independent} masses
and radii of \V6, is shown in Fig.~\ref{fig:v636cen_mr_y2}.  
For the primary component, the models predict an age of
2.6--3.2 Gyr, depending on the assumed \feh, whereas the
secondary is predicted to be more than two times older.
At the predicted age of the primary, the observed radius of the
secondary, which has been determined to an accuracy of
0.5\%, is about 10\% larger than that of the corresponding 
model. 
Mass-luminosity comparisons are shown in Fig.~\ref{fig:v636cen_ml_y2}.
Within the errors of the observed luminosities, the models predict 
identical - but lower - ages for the two components.

In conclusion, the secondary component of \V6\ is significantly larger 
and $\sim 400$~K cooler than predicted by models which 
adopt a mixing-length parameter matching the Sun. For the observed \feh, 
such models are $\sim 200$~K hotter than the primary component, indicating 
a discrepancy as well.

\subsection{Comparison with mixing-length ``tuned'' models}
\label{sec:models_lhp}

As mentioned in Sect.~\ref{sec:intro}, several authors have demonstrated
that models which adopt a reduced envelope convection mixing-length parameter
fit active late-type binary components better. 
Since \V6\ exhibits intrinsic variation and Ca\,\itwo\ H and K emission
(Fig.~\ref{fig:v636cen_ca}), and has relatively high rotational
velocities (Table~\ref{tab:v636_absdim}), compared to single field G-type stars,
we have investigated such models.

Models with reduced mixing-length parameters, including non-gray atmospheres
and an improved equation of state, have been published by 
Baraffe et al. (\cite{baraffe98})\footnote{{\scriptsize\tt ftp.ens-lyon.fr}},
but only for solar metallicity (up to 1.4 \Msun) and \feh\,$=-0.5$ (up to 1.0 \Msun).

Guided by a comparison between \V6\ and these models, we have calculated 
dedicated models for the observed masses and \feh, tuning the mixing length
parameter individually for the components. We have adopted the code by
Claret (\cite{c04}, \cite{c05}, \cite{c06}, \cite{c07}), which assumes
an enrichment law of $Y = 0.24 + 2.3Z$ 
together with the solar mixture by Grevesse \& Sauval (\cite{gs98}), 
leading to ($X$,$Y$,$Z$)$_{\sun}$ = (0.704,0.279,0.017).
The observed \feh\,$=-0.20$ then corresponds to ($X$,$Y$,$Z$) = (0.724,0.264,0.012).
We note that the envelope mixing length parameter needed to reproduce
the Sun is $l/H_p = 1.68$. 
\input v636cen_age.tex

\begin{figure}
\epsfxsize=90mm
\epsfbox{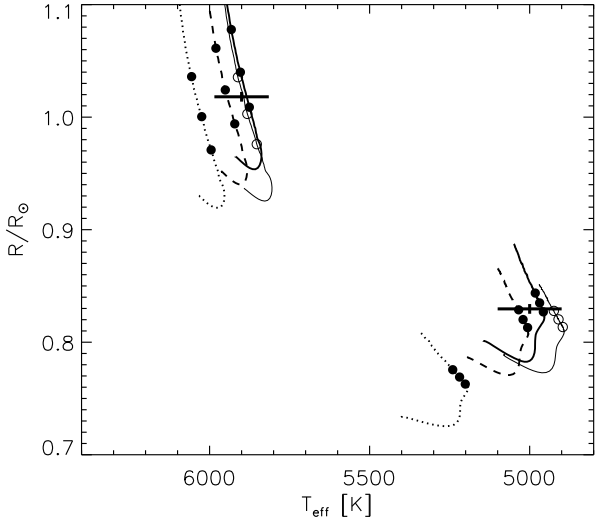}
\caption[]{\label{fig:v636cen_tr_claret}
\V6\ compared to Claret models for the observed masses, calculated
with different mixing length parameters. 
The four cases listed in Table~\ref{tab:v636cen_age} are shown.
The circles on the evolutionary tracks mark model ages of 
1.0, 2.0, and 3.0 Gyr.
Case 1 = full drawn, thick; Case 2 = dashed; Case 3 = full drawn,
thin, open circles; Case 4 = dotted.   
}
\end{figure}

\begin{figure}
\epsfxsize=90mm
\epsfbox{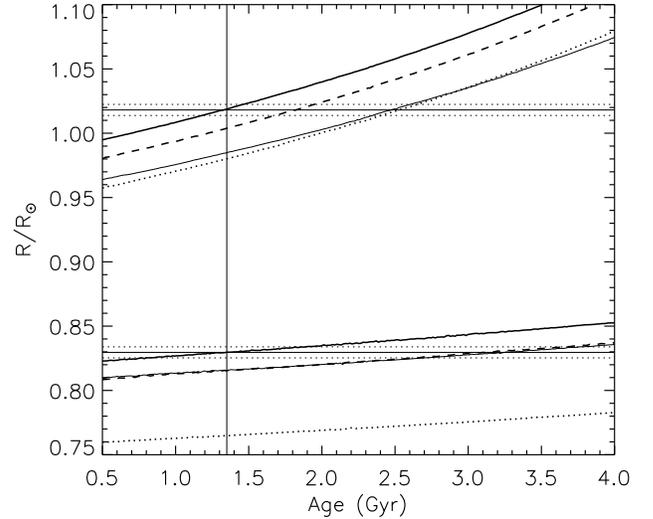}
\caption[]{\label{fig:v636cen_ar_claret}
\V6\ compared to Claret models.
The four cases listed in Table~\ref{tab:v636cen_age} are shown.
The curves illustrate model radii as function of age for the component
masses (upper: primary; lower: secondary).
Case 1 = full drawn, thick; Case 2 = dashed; Case 3 = full drawn,
thin; Case 4 = dotted.   
The horizontal full drawn lines mark the observed radii of the 
components with errors (dotted lines).
The vertical line marks the common age (1.35 Gyr) predicted by 
the case 1 models.
}
\end{figure}

Four different cases, listed in Table~\ref{tab:v636cen_age}, will be
discussed here.
Fig.~\ref{fig:v636cen_tr_claret} shows temperature-radius comparisons 
between \V6\ and models with different mixing length parameters, 
calculated for the component masses.
For the observed \feh, the temperature of the primary component is fitted 
well by the $l/H_p = 1.4$ model, whereas 1.0 is adequate for the secondary
(case 1). 
As seen from the age-radius diagram in Fig.~\ref{fig:v636cen_ar_claret}, 
these models also predict practically identical ages for the components. 
Changing the mixing length parameters by just $+0.1$ (case 2),
i.e. considering models which still fit the observed masses and 
temperatures within errors (Fig.~\ref{fig:v636cen_tr_claret}), 
lead to significantly higher ages, which are, however, no longer 
identical for the components.
To illustrate the effect of the abundance uncertainty $\pm 0.08$ dex, 
models for \feh\,$=-0.12$ have been included in the two figures (case 3).
For this abundance, mixing length parameters of about 1.6 (primary)
and 1.1 (secondary) are needed to fit the observed masses and
temperatures. Ages determined from the masses and radii become
even higher, and they differ for the components.
Finally, Claret models adopting a solar $l/H_p$ are, like the
$Y^2$ models, unable to fit \V6\ (case 4).

Table~\ref{tab:v636cen_age} summarizes the ages derived from the
different model cases. 
Effects from the mass uncertainties are negligible and have not been
included; see Fig.~\ref{fig:v636cen_tr_y2}.
In conclusion, only the case 1 models 
reproduce all observed properties of the components of \V6
well at a common age (1.35 Gyr). 
This implies that $l/H_p$ can be tuned fairly precisely, provided 
accurate dimensions and abundances are available.
We finally note, that according to the Claret models both components 
have pronounced convective envelopes, starting at $0.78\, R_p$ and
$0.73\, R_s$.

\subsection{Comparison with other binaries}
\label{sec:binaries}

\begin{figure}
\epsfxsize=100mm
\epsfbox{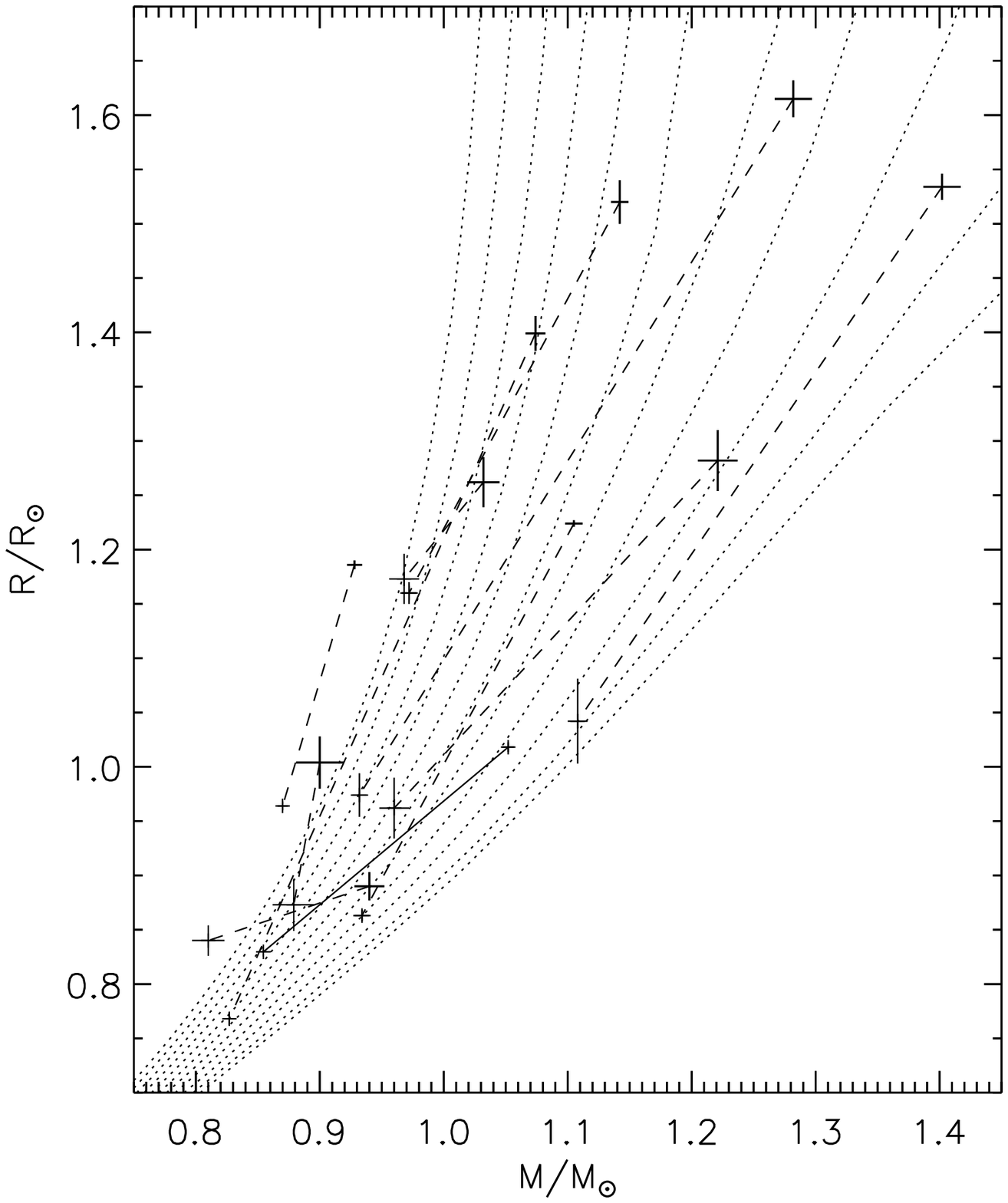}
\caption[]{\label{fig:debs}
Main-sequence binaries with one or both components in the
0.80--1.10 $M_{\sun}$ interval having well-established masses and radii 
(2\% or better); see Tabel~\ref{tab:systems}.
The components of V636\,Cen are connected by a solid line.
The dottes lines are $Y^2$ isochrones (0.5 Gyr and 1-10 Gyr, step 1 Gyr) 
for $Z=0.0181$, i.e. solar composition (Demarque et al. \cite{yale04}).
}
\end{figure}

\input systems.tex

Currently, nine other main sequence eclipsing binaries with 
at least one component in the 0.80--1.10 $M_{\sun}$ mass interval have 
well-established ($\approx2\%$ or better) masses and 
radii\footnote{see e.g. {\scriptsize\tt http://www.astro.keele.ac.uk/$\sim$jkt/}}
\footnote{Two additional binaries in this mass interval, the 
pre-main-sequence system V1174\,Ori (Stassun et al. \cite{stassun04}), 
and V432\,Aur, which has an evolved component (Siviero et al. \cite{sms04}), 
have not been been included in the comparison.}.
Furthermore, the A and B components of of $\alpha$\,Cen can be added, since
their masses, radii, temperatures, and abundances are known to high
accuracy (e.g. Eggenberger et al. \cite{eggenberger04}).

These 11 binaries are listed in Table~\ref{tab:systems} and shown in 
Fig.~\ref{fig:debs} together with $Y^2$ solar composition isochrones. 
We note, that \V6 and $\alpha$\,Cen are the only systems in this sample 
with \feh\ measured spectroscopically. 
With one exception, VZ\,Cep, the systems fall in two groups:

\begin{enumerate}
\item For \object{CG\,Cyg}, \object{CV\,Boo}, 
\object{ZZ\,UMa}, \object{FL\,Lyr}, and \object{V1061\,Cyg},
like for \V6, the $Y^2$ models predict higher ages for 
the secondary components than for the primary components, 
resulting from larger radii than predicted. Also, the
effective temperatures of the secondary components are lower
than predicted. Their orbital periods are short ($0\fd6$--$2\fd8$),
and the rotational velocities of their components are high (20--73 \kms).
Intrinsic variations are present, and some of the binaries are 
known, strong X-ray sources. 
Information on Ca\,\itwo\ emission is unfortunately scarce.
A few further systems with similar characteristics, but less accurate 
dimensions 
(\object{RT And},      
\object{BH Vir},      
\object{UV Leo}, and  
\object{UV Psc}),     
could be added to this group (Clausen et al. \cite{granada99}).

\item For \object{HS\,Aur}, \object{RW\,Lac}, \object{$\alpha$\,Cen}, and 
\object{V568\,Lyr} (V20 in \object{NGC\,6791}), the observed masses, 
radii, and temperatures agree much better with those predicted by the 
solar-scaled $Y^2$ models.
They have fairly long orbital periods ($9\fd9$--79.9 yr), 
and the rotational velocities of the components are below 6 \kms.
Popper et al. (\cite{dmp86}) saw no Ca\,\itwo\ emission 
but found evidence for intrinsic variability of HS\,Aur and/or 
the K0 comparison star used (HD49987). The other three binaries are
reported to be stable; no information on Ca\,\itwo\ emission is
available. None of the four is a known strong X-ray source.
According to models, RW\,Lac is somewhat metal deficient, 
and its age is about 11 Gyr (Lacy et at. \cite{ltcv05}). 
V568\,Lyr has an age of about 8 Gyr and its metal content is
about 2.5 times solar (Grundahl et al. \cite{gchf08}). 

\end{enumerate}

For the exception, \object{VZ\,Cep}, the observed masses and radii
are well reproduced by the $Y^2$ models, but the observed
temperature of the secondary component is 
about 250~K lower than predicted (Torres et al. \cite{tl09}).
Therefore, it is also less luminous than predicted, which is
in general not the case for the rest of the sample. 
VZ\,Cep is a strong X-ray source, and as for the binaries in group 1, 
the orbital period is short and the components rotate fast. 
On the other hand, no intrinsic variations are reported.

From a comparison of V1061\,Cyg and FL\,Lyr (group 1)
against RW\,Lac and HS\,Aur (group 2), Torres et al. (\cite{wt06})
identified chromospheric activity as the likely cause of the
radius and temperature discrepancies.
Our sample strengthens this suggestion, not at least
through a comparison between \V6\ and V568\,Lyr, whose components have
nearly identical masses.

\subsection{Mixing-length calibrations}
\label{sec:mix}

It is worth linking our sample to the ongoing discussion on the 
mixing-length parameter for stars in the 0.7-1.1 $M_{\sun}$ range, where stars 
have convective envelopes but are not fully convective.
From 2D radiation hydrodynamics (RHD) calculations, Ludwig et al. (\cite{hgl99})
calibrated the mixing-length parameter for solar-type stars ($T_{\rm eff}$ 4300--7100~K;
$\log g$ 2.54-4.74) and found a moderate but significant variation of 
$l/H_p$. On the main sequence, a decrease with increasing temperature, and 
thereby mass, from about 1.8 to about 1.3 was obtained.
On the other hand, Yildiz et al. (\cite{yildiz06}) and Yildiz (\cite{yildiz08})
found a significant increase with mass from model fits to four Hyades binaries,
FL\,Lyr, and \object{V442\,Cyg}.
As an illustration, $l/H_P$ values calculated from the fitting functions
by these authors are listed in Table~\ref{tab:lhp}; no scaling, e.g. to
the Sun, has been done.
Finally, including seismic constraints, several studies of $\alpha$\,Cen, 
most recently by Eggenberger et al. (\cite{eggenberger04}) and Yildiz (\cite{yildiz07}), 
find that $l/H_p$ is higher for the less massive B component
than for the A component.
Eggenberger et al., who use the seismological data by Carrier \& Bourban
(\cite{cb03}), obtain 1.83 (A) and 1.97 (B), and Yildiz, who include the
data by Kjeldsen et al. (\cite{hk05}), get 1.64 (A) and 2.10 (B).
 
\input lhp.tex

The fact that our sample can be divided in two different groups seems to 
suggest that the behaviour of the mixing length parameter may be governed 
by two different effects:
\begin{itemize} 
\item a slight decrease of $l/H_p$ with increasing temperature/mass for 
inactive main sequence stars, as predicted by Ludwig et al. (\cite{hgl99}) 
and found for $\alpha$\,Cen and possibly also V568\,Lyr
(Grundahl et al. \cite{gchf08}), and 
\item a decrease of $l/H_p$ for active stars compared to inactive ones
of similar mass, as seen for group 2.
\end{itemize}

\noindent
If this is the case, the result by Yildiz et al. (\cite{yildiz06}) and Yildiz 
(\cite{yildiz08}) might then be questioned, since at least two of the binaries 
in their samples, the Hyades binary \object{V818\,Tau} and FL\,Lyr, have
active component(s).

It would be of interest to extend the systematic study of the evolution
of low-mass stars and brown dwarf eclipsing binaries by
Chabrier et al. (\cite{cgb07}) to higher masses. For masses up to
0.8 \Msun, they found that the effects of decreasing $l/H_p$, i.e. reducing 
the convective efficiency, increases with mass, leading to lower effective 
temperatures, luminosities, central temperatures, and nuclear energy 
production. In parallel, magnetic spots covering a significant fraction 
of the stellar surfaces cause similar changes. 
They adopted a phenomenological approach, and more complete models,
which directly couple rotation, convection, and dynamo magnetic fields are
also desirable; see e.g. D'Antona et al. (\cite{da00}) for pre-main-sequence
models. 

\section{Summary and conclusions}
\label{sec:sum}

From state-of-the-art observations and analyses, precise (0.5\%) absolute 
dimensions have been obtained for the totally eclipsing solar-type system \V6. 
A detailed spectroscopic analysis yields a Fe abundance relative to the Sun
of \feh\,$=-0.20\pm0.08$ and similar relative abundances for Si, Ca, Ti, V, Cr, Co, 
and Ni.

The 0.85 \Msun\ secondary component is found to be moderately active with 
starspots and Ca\,\itwo\ H and K emission, and the 1.05 \Msun\ primary shows 
signs of activity as well, but at a much lower level.
Apsidal motion ($U = 5\,270 \pm 335$ yr) with a 40\% relativistic contribution 
has been detected for the eccentric orbit ($e = 0.135 \pm 0.001$), and the
inferred mean central density concentration coefficient, 
log($k_2$) = $-1.61 \pm 0.05$, agrees marginally with model predictions.  
The measured rotational velocities, $13.0 \pm 0.2$ (primary) and 
$11.2 \pm 0.5$ (secondary) \kms, are in remarkable agreement with the 
theoretically predicted pseudo-synchronous velocities, but are about 15\% 
lower than the periastron values.

We have shown that stellar models with solar-scaled mixing length parameters 
are unable to match the components of \V6 at identical ages.
At the age of the primary, the secondary component is $\sim 10\%$ larger 
than predicted. Also, the components are $\sim 200$~K (primary) and
$\sim 400$~K (secondary) cooler than predicted. 

However, models adopting significantly lower mixing-length parameters can 
remove these discrepancies.
For the observed \feh, Claret models for $l/H_p = 1.4$ (primary) and 1.0 
(secondary) reproduce the properties of the components of \V6\ well for a 
common age of 1.35 Gyr.

Currently, 10 other solar-type binaries have well-estqblished dimensions,
although spectroscopic abundance results are lacking for most of
them; see Fig.~\ref{fig:debs} and Table~\ref{tab:systems}. 
With one exception, VZ\,Cep, they fall in two groups:
Four long period, slowly rotating inactive systems,
which seem to be fitted reasonably well by solar-scaled models.
And five systems, which exhibit intrinsic variation, and have short orbital 
periods and high rotational velocities. Like \V6, they can not be reproduced
by solar-scaled models.
Therefore, the sample strengthens the suggestion by Torres et al.
(\cite{wt06}), based on fewer systems, that chromospheric activity, 
and the associated changes of the convective envelope, is the 
likely cause of the radius and temperature discrepancies. 
The strongest case is perhaps a comparison between \V6\ (active) and 
the newly studied V568\,Lyr in NGC\,6791 (inactive), whose components have nearly 
identical masses.

Finally, the clear division of our sample in two groups might
suggest that two different effects influence the mixing length
parameter for 0.7--1.1 $M_{\sun}$ main sequence stars:
a) a slight decrease of $l/H_p$ with increasing temperature/mass for inactive 
main sequence stars, as found by e.g. Ludwig et al. (\cite{hgl99}) and seen for
$\alpha$\,Cen and possibly also V568\,Lyr, 
and b) a decrease of $l/H_p$ for active stars compared to 
inactive ones of similar mass.

More well-studied eclipsing binaries, as well as more complete stellar models, 
are needed to fully understand and describe
envelope convection in terms of effective temperature,
gravity, mass, age, rotation, dynamo magnetic field, activity level etc.
We are presently studying several systems with main sequence components 
in the 0.5--1.2 $M_{\sun}$ range, including
\object{V1123\,Tau},  
\object{V963\,Cen},   
\object{AL\,Dor},     
\object{QR\,Hya},     
\object{KY\,Hya},     
\object{AL\,Ari},     
\object{UW\,LMi},     
\object{EW\,Ori}, and 
\object{NY\,Hya};     
see also Clausen et al. (\cite{jvcetal01}).
They exhibit various levels of activity, from none to modest, 
and should therefore provide valuable new insight into this matter.
Spectroscopic abundance determinations will be part of the analyses.

\begin{acknowledgements}
We thank M. Imbert and L. Pr\'evot for the use of their CORAVEL
observations, and P. Figueira for extracting the radial velocity data from 
the CORAVEL database. 
We are grateful to the referee, G. Torres, for a prompt and very
helpful report.
The projects "Stellar structure and evolution - new challenges from
ground and space observations" and "Stars: Central engines of the evolution
of the Universe", carried out at Copenhagen University
and Aarhus University, are supported by the Danish National Science
Research Council.
Furthermore, this investigation was supported by the 
Spanish Science Research Council. 
JA and BN acknowledge support from the Carlsberg Foundation.
HB was supported by the Australian Research Council.
The following internet-based resources were used in research for
this paper: 
the NASA Astrophysics Data System; 
the SIMBAD database and the ViziR service operated by CDS, Strasbourg, France; 
the VALD database made available through the Institute of Astronomy,
Vienna, Austria.
\end{acknowledgements}

{}

\listofobjects

\appendix

\section{Radial velocity observations}
\label{sec:rvtab}
\input v636_rv.tex

\end{document}

%% file: 8788_ebop_vh.tex
\begin{table}
\caption[]{\label{tab:8788_ebop_vh}
Photometric solutions for \V6 from JKTEBOP analyses of the 1987 and 1988
light curve observations (426 points per band). 
}
\begin{center}
\begin{tabular}{lrrrr} \hline
\hline\noalign{\smallskip}
                     &     $y$    &       $b$  &       $v$  &   $u$\\                 
\noalign{\smallskip}
\hline
\noalign{\smallskip}
$i$ \, (\degr)       &  89.65     &   89.62    &   89.62    &  89.63\vspace{-0.8mm}\\ 
                     & $\pm 7$    &  $\pm 7$   &  $\pm 8$   & $\pm 9$\\               

$r_p$                &  0.0746    &   0.0748   &   0.0750   &  0.0751\vspace{-0.8mm}\\
                     &  $\pm 5$   &   $\pm 5$  &   $\pm 6$  &  $\pm 6$\\              

$r_s$                &  0.0613    &   0.0614   &   0.0613   &  0.0614\vspace{-0.8mm}\\
                     &  $\pm 3$   &   $\pm 3$  &   $\pm 3$  &  $\pm 3$\\
                                                                                       
$k$                  &   0.822    &   0.822    &   0.818    &   0.817\vspace{-0.8mm}\\ 
                     &  $\pm 5$   &  $\pm 5$   &  $\pm 6$   & $\pm  6$\\              

$r_p + r_s$          &  0.1359    &   0.1362   &   0.1363   &  0.1365\vspace{-0.8mm}\\
                     &  $\pm 6$   &   $\pm 5$  &   $\pm 7$  &  $\pm 7$\\
                                                                                      
$u_p$                &  0.59      &   0.68     &   0.77     &  0.81\\   

$u_s$                &  0.73      &   0.84     &   0.95     &  0.99\\

$y_p$                &  0.37      &   0.43     &   0.49     &  0.58\\

$y_s$                &  0.44      &   0.51     &   0.59     &  0.70\\

$J_s/J_p$            &  0.417     &    0.370   &   0.286    &  0.236\vspace{-0.8mm}\\
                     &  $\pm10$   &   $\pm 9$  &   $\pm 8$  &  $\pm 7$\\

$L_s/L_p$            &  0.265     &    0.233   &   0.176    &  0.144\vspace{-0.8mm}\\
                     & $\pm 3$    &   $\pm 3$  &  $\pm 3$   & $\pm 3$\\

$\sigma$ \, (mmag.)  &  11.3      &    11.4    &   13.2     &  14.5\\
\noalign{\smallskip}            
\hline
\end{tabular}            
\end{center}            
\textsc{Note 1:}
Orbital parameters of $e = 0.1348$ and $\omega = 281\fdg60$ were assumed
(Table~\ref{tab:v636_orb}), and linear limb darkening coefficients by Van 
Hamme (\cite{vh93}) were adopted.

\textsc{Note 2:}
The errors quoted for the free parameters are realistic values determined from 
1\,000 Monte Carlo simulations. $e$ and $\omega$ were perturbed corresponding
to their uncertainties, and the limb darkening coefficients within $\pm 0.05$.
\end{table}

%% file: 1987_wd_vh.tex
\begin{table}
\caption[]{\label{tab:1987_wd_vh}
WD solutions from the 1987 light curve observations (165 per band).
}
\begin{center}
\begin{tabular}{lrrrr} \hline
\hline\noalign{\smallskip}
                     &     $y$    &       $b$  &       $v$  &   $u$\\                   
\hline\noalign{\smallskip}            
$i$ \, (\degr)       &  89.80     &   89.63    &   89.83    &  89.81\vspace{-0.8mm}\\   
                     & $\pm 3$    &  $\pm 2$   &  $\pm 2$   & $\pm 4$\\                 

$\Omega_p$           &  14.504    &   14.465   &   14.395   &  14.397\vspace{-0.8mm}\\  
                     &  $\pm22$   &   $\pm33$  &   $\pm38$  &  $\pm49$\\                

$\Omega_s$           &  14.738    &   14.666   &   14.725   &  14.720\vspace{-0.8mm}\\  
                     &  $\pm17$   &   $\pm34$  &   $\pm40$  &  $\pm56$\\                

$r_p$$^{\mathrm{a}}$    &  0.0737    &   0.0739   &   0.0743   &  0.0743\\             
                                                                                        
$r_s$$^{\mathrm{a}}$     &  0.0601    &   0.0604   &   0.0601   &  0.0602\\            
                                                                                      
$k$                  &  0.815     &   0.817    &   0.809    &  0.810\\               
                                                                                    
$r_p + r_s$          &  0.1338    &   0.1343   &   0.1344   &  0.1345\\            
                                                                                  
$T_s$ \, (K)         &  4965      &   4984     &   4916     &  4962\vspace{-0.8mm}\\
                     &$\pm 6$     & $\pm 5$    &$\pm  7$    &$\pm 7$\\

$L_s/L_p$            &  0.286     &    0.256   &   0.199    &  0.172  \\

$LO_1$ \, (\degr)    & 12.2       &   15.0     &   17.2     &  31.8\vspace{-0.8mm}\\
                     &$\pm 3.7$   & $\pm 3.2$  & $\pm 3.3$  & $\pm 6.6$\\

$AR_1$ \, (\degr)    & 23.9       &   25.7     &   29.6     &  31.8\vspace{-0.8mm}\\
                     &$\pm 1.0$   & $\pm 0.8$  & $\pm 1.0$  & $\pm 1.2$\\

$LO_2$ \, (\degr)    & 233.0      &   235.3    &   233.8    &  255.6\vspace{-0.8mm}\\
                     &$\pm 7.6$   & $\pm 3.7$  & $\pm 3.5$  & $\pm 10.1$\\

$AR_2$ \, (\degr)    & 17.7       &   19.3     &   22.5     &   25.7\vspace{-0.8mm}\\
                     &$\pm 1.4$   & $\pm 0.8$  & $\pm 1.2$  &  $\pm 1.2$\\

$\sigma$ \, (mmag.)  &  6.2       &  6.2       &   7.7      &  9.1     \\
\noalign{\smallskip}            
\hline
\end{tabular}            
\begin{list}{}{}
\item[$^{\mathrm{a}}$] relative volume radius
\end{list}
\end{center}            
\textsc{Note 1:}
$T_p$ was assumed to be 5900 K; for orbital parameters and limb darkening
coefficients see Table~\ref{tab:8788_ebop_vh}.

\textsc{Note 2:}
The errors quoted for the free parameters are the $formal$ probable errors 
determined from the iterative least squares solution procedure.
\end{table}                       

%% file: 1988_wd_vh.tex
\begin{table}
\caption[]{\label{tab:1988_wd_vh}
WD solutions from the 1988 light curve observations (261 per band).
See Table~\ref{tab:1987_wd_vh}.
}
\begin{center}
\begin{tabular}{lrrrr} \hline
\hline\noalign{\smallskip}
                     &     $y$    &       $b$  &       $v$  &   $u$\\ 
\hline\noalign{\smallskip}            
$i$ \, (\degr)       &  89.56     &   89.55    &   89.56    &  89.40\vspace{-0.8mm}\\
                     & $\pm 3$    &  $\pm 3$   &  $\pm 3$   & $\pm 2$\\               

$\Omega_p$           &  14.495    &   14.470   &   14.455   &  14.477\vspace{-0.8mm}\\
                     &  $\pm23$   &   $\pm20$  &   $\pm25$  &  $\pm27$\\              

$\Omega_s$           &  14.656    &   14.639   &   14.707   &  14.623\vspace{-0.8mm}\\ 
                     &  $\pm27$   &   $\pm25$  &   $\pm14$  &  $\pm19$\\               

$r_p$                &  0.0738    &   0.0739   &   0.0740   &  0.0739\\               
                                                                                      
$r_s$                &  0.0604    &   0.0605   &   0.0602   &  0.0606\\              
                                                                                     
$k$                  &  0.818     &   0.819    &   0.814    &  0.820\\               
                                                                                    
$r_p + r_s$          &  0.1342    &   0.1344   &   0.1342   &  0.1345\\             
                                                                                   
$T_s$ \, (K)         &  4940      &   4959     &   4897     &  4914\vspace{-0.8mm}\\
                     &$\pm 4$     & $\pm 4$    &$\pm  4$    &$\pm 6$\\

$L_s/L_p$            &  0.282     &    0.250   &   0.195    &  0.163  \\

$LO_1$ \, (\degr)    & 107.9      &   118.8    &   116.5    & 122.0\vspace{-0.8mm}\\
                     &$\pm 1.6$   & $\pm 1.2$  & $\pm 1.4$  & $\pm 2.6$\\

$AR_1$ \, (\degr)    & 25.9       &   28.7     &   31.8     &  31.1\vspace{-0.8mm}\\
                     &$\pm 0.4$   & $\pm 0.4$  & $\pm 0.4$  & $\pm 0.6$\\

$LO_2$ \, (\degr)    & 202.4      &   212.1    &   209.9    &  214.9\vspace{-0.8mm}\\
                     &$\pm 1.7$   & $\pm 1.6$  & $\pm 1.6$  & $\pm  2.4$\\

$AR_2$ \, (\degr)    & 25.7       &   23.8     &   28.6     &   29.9\vspace{-0.8mm}\\
                     &$\pm 0.6$   & $\pm 0.7$  & $\pm 0.7$  &  $\pm 0.9$\\

$\sigma$ \, (mmag.)  &  6.5       &  5.9       &   6.8      &  9.7     \\
\noalign{\smallskip}            
\hline
\end{tabular}            
\end{center}            
\end{table}

%% file: v636_phel.tex
\begin{table}            
\caption[]{\label{tab:v636_phel}
Adopted photometric elements for \V6.
}
\begin{center}             
\begin{tabular}{ll} \hline            
\hline\noalign{\smallskip}             
$i$              & $89{\fdg}65 \pm 0{\fdg}10$ \\
$r_p$            & $0.0740 \pm 0.0003$ \\       
$r_s$            & $0.0603 \pm 0.0003$ \\       
\noalign{\smallskip}             
\hline
\end{tabular}             
\begin{tabular}{lrrrr}             
\noalign{\smallskip}             
                 & $y$    & $b$    & $v$   & $u$  \\           
\noalign{\smallskip}             
$L_s/L_p$        & 0.269  & 0.237  & 0.178 & 0.152\vspace{-0.8mm} \\  
                 &$\pm15$ &$\pm15$ &$\pm15$&$\pm15$\\  
\noalign{\smallskip}             
\hline             
\end{tabular}             
\end{center}            
\textsc{Note:} The individual luminosity ratios are based
on the mean stellar and orbital parameters and include the
average activity level (spots) in 1987 and 1988.
\end{table}

%% file: v636_orbit.tex
\begin{table*}
\caption[]{\label{tab:v636_orb}
Spectroscopic orbital elements of \V6. 
}
\begin{center}
\begin{tabular}{lrrrr} \hline 
\hline\noalign{\smallskip}
                            &  Primary            & Secondary           & Primary+Secondary\\
                            &                     &                     & Adopted solution\\
\hline\noalign{\smallskip}
$K_p$ \,(\kms)              &$ 73.54 \pm 0.07$    &                     &$73.53  \pm 0.10$\\
$K_s$ \,(\kms)              &                     &$90.48  \pm 0.25$    &$90.51  \pm 0.19$\vspace{1.0mm}\\

$\gamma$ \,(\kms)           &$-14.23 \pm 0.07$    &$-14.31 \pm 0.24$    &$-14.25  \pm 0.07$\vspace{1.0mm} \\

$e$                         &$0.1353  \pm 0.0010$ &$0.1339 \pm 0.0029$  &$0.1348 \pm 0.0012$\vspace{1.0mm}\\    

$\omega$ \,(\degr)          &$281.57 \pm 0.58$    &$281.23 \pm 1.78$    &$281.60  \pm 0.63$\vspace{1.0mm} \\

$T$-2\,400\,000$^{\mathrm{a}}$ &$47223.1225\pm0.0066$&$47223.1200\pm0.0199$&$47223.1233 \pm 0.0072$\vspace{1.0mm}\\

$\sigma_p$$^{\mathrm{b}}$ \,(\kms)         & 0.33                &                     & 0.42\\
$\sigma_s$$^{\mathrm{b}}$ \,(\kms)         &                     & 1.23                & 0.79\vspace{1.0mm}\\
\noalign{\smallskip}
\hline
\end{tabular}
\begin{list}{}{}
\item[$^{\mathrm{a}}$] Time of periastron
\item[$^{\mathrm{b}}$] Standard deviation of a single radial velocity
\end{list}
\end{center}
\textsc{Note:}
An orbital period of $4\fd2839423$ has been assumed (CVG08).
\end{table*}

%% file: feros.tex
\begin{table}
\caption[]{\label{tab:feros}
Log of the FEROS observations of \V6.
}
\begin{minipage}{\columnwidth}
\centering
\renewcommand{\footnoterule}{}  
\begin{tabular}{ccrr}
\hline
\hline\noalign{\smallskip}
HJD$-$2\,400\,000\footnote{Refers to mid-exposure}  & phase &t$_{exp}$\footnote{Exposure time in seconds} & S/N\footnote{Signal-to-noise ratio measured around 6070 {\AA}}\\
\noalign{\smallskip}
\hline
\noalign{\smallskip}
51210.8718         &  0.4017 & 2100 &  160   \\   
51211.8484         &  0.6296 & 2400 &  200   \\   
\hline
\end{tabular}
\end{minipage}
\end{table}

%% file: v636cen_abund.tex
\begin{table}
\caption[]{\label{tab:v636cen_abund}
Abundances ($[\mathrm{El./H}]$) for the primary and secondary
components of \V6 determined from the two FEROS spectra.
}
\begin{center}
\begin{tabular}{lllrllr} \hline
\hline\noalign{\smallskip}
             &  \multicolumn{3}{c}{Primary} & \multicolumn{3}{c}{Secondary} \\
Ion          &[El./H]&  rms& N$_t$/N$_l$&[El./H]& rms & N$_t$/N$_l$  \\
\noalign{\smallskip}
\hline
Si\ione\     &$-0.17$& 0.12& 15/15 &$-0.26$& 0.15&  3/3 \\  
Ca\ione\     &$-0.15$& 0.11& 11/9  &$-0.10$& 0.23& 12/9 \\  
Ti\ione\     &$-0.19$& 0.15&  5/5  &$-0.34$& 0.16&  8/6 \\  
Ti\itwo\     &$-0.22$& 0.10&  6/4  &       &     &    \\  
V\ione\      &$-0.27$& 0.04&  6/4  &$-0.15$& 0.17& 13/12 \\
Cr\ione\     &$-0.19$& 0.15&  6/4  &       &     &    \\
Cr\itwo\     &$-0.05$& 0.16&  6/5  &       &     &    \\
Fe\ione\     &$-0.19$& 0.09&136/110&$-0.11$& 0.19& 48/46 \\
Fe\itwo\     &$-0.20$& 0.08& 16/12 &       &     &    \\ 
Co\ione\     &$-0.15$& 0.15&  7/5  &       &     &    \\
Ni\ione\     &$-0.22$& 0.13& 39/25 &$-0.19$& 0.17& 12/9 \\
\noalign{\smallskip}
\hline
\end{tabular}            
\end{center}
\textsc{Note:}
N$_t$ is the total number of lines used per ion, and N$_l$ is the
number of different lines used per ion.
Ions with at least 3 lines measured are included.
\end{table}

%% file: v636_absdim.tex
\begin{table}   
\caption[]{\label{tab:v636_absdim}
Astrophysical data for \V6.
}
\begin{minipage}{\columnwidth}
\begin{center}    
\begin{tabular}{lrr} \hline    
\noalign{\smallskip}    
\hline    
\noalign{\smallskip}    
                     &    Primary       &    Secondary      \\ 
\noalign{\smallskip}    
\hline    
\noalign{\smallskip}    
Absolute dimensions:          &                   &                 
 \\ 
$M/M_{\sun}$                  &$1.052 \pm 0.005$  &$0.854 \pm 0.003$
\\ 
$R/R_{\sun}$                  &$1.018 \pm 0.004$  &$0.830 \pm 0.004$ 
\\ 
$\log g$ (cgs)                & $4.444 \pm 0.004$ & $4.532 \pm 0.005$
\\
$v \sin i$$^{\mathrm{a}}$ (\kms)   & $13.0 \pm 0.2$    & $11.1 \pm 0.5$      
\\ 
$v_{sync}$$^{\mathrm{b}}$ (\kms)   & $12.0 \pm 0.1$    & $ 9.8 \pm 0.1$ 
\\ 
$v_{psync}$$^{\mathrm{c}}$ (\kms)  & $13.3 \pm 0.1$    & $10.9 \pm 0.1$ 
\\ 
$v_{peri}$$^{\mathrm{d}}$ (\kms)   & $15.9 \pm 0.1$    & $13.0 \pm 0.1$ 
\\ 
 & & \\ 
Photometric data:&                       &                         \\ 
$V$$^{\mathrm{e}}$     &         $8.963 \pm 0.017$  &        $10.389 \pm 0.049$\\  
$(b-y)$$^{\mathrm{e}}$ &         $0.382 \pm 0.004$  &        $ 0.520 \pm 0.009$\\
$m_1$$^{\mathrm{e}}$   &         $0.179 \pm 0.007$  &        $ 0.357 \pm 0.017$\\
$c_1$$^{\mathrm{e}}$   &         $0.315 \pm 0.018$  &        $ 0.169 \pm 0.020$\\
$E(b-y)$    & \multicolumn{2}{c}   {$0.005 \pm 0.008$}  \\
$T_{\mbox{\scriptsize eff}}\,$       &  $5900 \pm 85$     &   $5000 \pm 100$ \\
$M_{\mbox{\scriptsize bol}}\,$      &  $4.61  \pm 0.06$  &   $5.78  \pm 0.09$ \\
$\log L/L_{\sun}$ & $0.05 \pm 0.03$ &    $-0.41 \pm 0.04$ \\
$BC$              & $-0.06$         &    $-0.30$ \\
$M_V$ &             $ 4.67 \pm 0.06$&   $ 6.08 \pm 0.09$ \\
$V-M_V$          &$ 4.29  \pm 0.07 $& $ 4.31  \pm 0.10 $ \\  
$V_0-M_V$          &$ 4.27  \pm 0.08 $& $ 4.29  \pm 0.11 $ \\   
Distance \, (pc) &$  71.4 \pm  2.5 $& $  72.2 \pm  3.6 $ \\
           &              &               \\
Abundance: &              &               \\
\feh\                          & \multicolumn{2}{c}   {$-0.20 \pm 0.08$} \\
\noalign{\smallskip}            
\hline
\noalign{\smallskip}
\end{tabular}            
\begin{list}{}{}
\item[$^{\mathrm{a}}$] Observed rotational velocity
\item[$^{\mathrm{b}}$] Equatorial velocity for synchronous rotation
\item[$^{\mathrm{c}}$] Equatorial velocity for pseudo-synchronous rotation
\item[$^{\mathrm{d}}$] Refers to periastron velocity
\item[$^{\mathrm{e}}$] Not corrected for interstellar absorption/reddening
\end{list}
\end{center}            
\textsc{Note:} 
Bolometric corrections ($BC$) by Flower (\cite{flower96}) have been 
assumed, together with 
$T_{eff\sun} = 5780$ K, $BC_{\sun} = -0.08$, and $M_{bol\sun} = 4.74$.
\end{minipage}
\end{table}

%% file: v636cen_age.tex
\begin{table}
\caption[]{\label{tab:v636cen_age}
Ages (Gyr) for the primary ($p$) and secondary ($s$) components
of \V6\ determined from comparison between their masses and radii
and Claret models for different mixing length parameters and abundances. 
}
\begin{center}
\begin{tabular}{cccccc}
\hline
\hline\noalign{\smallskip}
     & \feh\ & $l/H_p(p)$ & $l/H_p(s)$ & Age(p)  & Age(s) \\
\noalign{\smallskip}
\hline
\noalign{\smallskip}
1 & -0.20  & 1.40      &  1.00      & $1.33 \pm  0.13$ & $1.37 \pm 0.55$ \\  
2 & -0.20  & 1.50      &  1.10      & $1.82 \pm  0.13$ & $3.10 \pm 0.55$ \\ 
3 & -0.12  & 1.60      &  1.10      & $2.46 \pm  0.12$ & $3.25 \pm 0.53$ \\ 
4 & -0.20  & 1.68      &  1.68      & $2.40 \pm  0.13$ &                 \\ 
\hline
\end{tabular}
\end{center}
\textsc{Note:}
Ages in Gyr.
Except for those with the solar $l/H_p = 1.68$ (case 4), the models fit
the observed effective temperatures within errors. 
See text for details.
\end{table}

%% file: systems.tex
\begin{table*}
\caption[]{\label{tab:systems}
Properties of well-studied main sequence binaries (e.g. footnote 9) with at least one 
component in the 0.80--1.10 $M_{\sun}$ mass interval.
}
\begin{minipage}{\textwidth}
\centering
\renewcommand{\footnoterule}{}  
\begin{tabular}{lrrrrlllr} \hline
\hline\noalign{\smallskip}
System       &Period &  $M/M_{\sun}$   & $R/R_{\sun}$    &$\log L/L_{\sun}$&
$v \sin i$\footnote{If measured rotational velocities are not available, synchronous values (s) are given}
  & Variability    &Ca\,\itwo & $\log L_X$\footnote{X-ray luminosities are based on Voges 
et al. (\cite{rosat99,rosat00}) and Motch et al. (\cite{motch97})}      \\
             &       &                 &                 &                 &   $(\kms)$ & (light curve)  &emission  & (cgs)\\
\hline\noalign{\smallskip}
Group 1:     &          &                 &                 &                &            &           &           &        \\
CG\,Cyg      & $0\fd63$ & $0.940\pm0.012$ & $0.890\pm0.013$ & $-0.26\pm0.06$ & 71(s)      & YES       &           &$<29.4$ \\ 
             &          & $0.810\pm0.013$ & $0.840\pm0.014$ & $-0.51\pm0.03$ & 67(s)      &           &           &        \\
CV\,Boo      & $ 0\fd85$& $1.032\pm0.013$ & $1.262\pm0.023$ & $ 0.20\pm0.05$ &$73\pm10$   & YES       &           & 30.7   \\ 
             &          & $0.968\pm0.012$ & $1.173\pm0.023$ & $ 0.11\pm0.05$ &$67\pm10$   &           &           &        \\
V636\,Cen    & $4\fd28$ & $1.052\pm0.005$ & $1.018\pm0.004$ & $ 0.05\pm0.03$ &$13.0\pm0.2$& YES       &  YES      & 30.0   \\ 
             &          & $0.854\pm0.003$ & $0.830\pm0.004$ & $-0.41\pm0.04$ &$11.1\pm0.5$&           &  YES      &        \\
ZZ\,UMa      & $2\fd80$ & $1.142\pm0.007$ & $1.520\pm0.020$ & $ 0.42\pm0.02$ & 33(s)      & YES       &           &$<29.6$ \\ 
             &          & $0.972\pm0.007$ & $1.160\pm0.010$ & $-0.03\pm0.03$ & 26(s)      &           &           &        \\
FL\,Lyr      & $2\fd18$ & $1.221\pm0.016$ & $1.282\pm0.028$ & $ 0.32\pm0.03$ &$30\pm2$    & ?/YES     &  NO       & 30.2   \\ 
             &          & $0.960\pm0.012$ & $0.962\pm0.028$ & $-0.18\pm0.04$ &$25\pm2$    &           &  NO       &        \\
V1061\,Cyg   & $2\fd35$ & $1.282\pm0.015$ & $1.615\pm0.017$ & $ 0.53\pm0.03$ &$36\pm2$    & ?/YES     &           & 30.1   \\ 
             &          & $0.932\pm0.007$ & $0.974\pm0.020$ & $-0.17\pm0.05$ &$20\pm3$    &           &           &        \\
\hline\noalign{\smallskip}
Group 2:     &          &                 &                 &                &            &           &           &        \\
HS\,Aur      & $9\fd85$ & $0.900\pm0.019$ & $1.004\pm0.024$ & $-0.13\pm0.03$ & 5(s)       & ?/YES     &  NO       &$<29.3$ \\ 
             &          & $0.879\pm0.017$ & $0.873\pm0.024$ & $-0.30\pm0.03$ & 5(s)       &           &  NO       &        \\
RW\,Lac      &$10\fd37$ & $0.928\pm0.006$ & $1.186\pm0.004$ & $ 0.14\pm0.03$ &$2\pm2$     & NO        &           &$<29.6$ \\
             &          & $0.870\pm0.004$ & $0.964\pm0.004$ & $-0.10\pm0.05$ &$0\pm2$     &           &           &        \\
$\alpha$\,Cen&79.9 yr   & $1.050\pm0.007$ & $1.224\pm0.003$ & $ 0.18\pm0.01$ & $2.7\pm0.7$& NO        &           &        \\   
             &          & $0.934\pm0.007$ & $0.863\pm0.005$ & $-0.30\pm0.02$ & $1.1\pm0.8$&           &           &        \\
V568\,Lyr    &$14\fd47$ & $1.074\pm0.008$ & $1.399\pm0.016$ & $ 0.26\pm0.03$ & 5(s)       & NO        &           &        \\
              &          & $0.827\pm0.004$ & $0.768\pm0.006$ & $-0.52\pm0.04$ & 3(s)       &           &           &        \\
\hline\noalign{\smallskip}
Exception:   &          &                 &                 &                &            &           &           &        \\
VZ\,Cep      & $1\fd18$ & $1.402\pm0.015$ & $1.534\pm0.012$ & $ 0.63\pm0.04$ &$57\pm3$    & NO        &           & 30.6   \\ 
             &          & $1.108\pm0.008$ & $1.042\pm0.039$ & $ 0.03\pm0.05$ &$50\pm10$   &           &           &        \\
\noalign{\smallskip}
\hline
\end{tabular}
\end{minipage}
\end{table*}

%% file: lhp.tex
\begin{table}
\caption[]{\label{tab:lhp}
Mixing-length parameters ($l/H_p$) for the binary components 
(Table~\ref{tab:systems})
calculated from the temperature-gravity fitting function by Ludwig et al. 
(\cite{hgl99}) and the mass fitting formula by Yildiz et al. (\cite{yildiz06}).
}
\begin{minipage}{\columnwidth}
\centering
\renewcommand{\footnoterule}{}  
\begin{tabular}{lcccc} \hline
\hline\noalign{\smallskip}
  & \multicolumn{2}{c} {Ludwig et al.} & \multicolumn{2}{c} {Yildiz et al.} \\
             & Pri. &   Sec.  &   Pri. &  Sec.\\
\hline\noalign{\smallskip}
Sun\footnote{For comparison, the values calculated for the Sun are included}
& \multicolumn{2}{c} {1.59} & \multicolumn{2}{c} {1.91} \\
\hline\noalign{\smallskip}
Group 1:     &      &         &        &      \\
CG\,Cyg      & 1.61 &   1.68  &   1.79 &  1.33 \\
CV\,Boo      & 1.58 &   1.59  &   1.96 &  1.85 \\
V636\,Cen    & 1.58 &   1.63  &   1.99 &  1.54 \\
ZZ\,UMa      & 1.55 &   1.61  &   2.11 &  1.86 \\
FL\,Lyr      & 1.54 &   1.61  &   2.19 &  1.83 \\ 
V1061\,Cyg   & 1.51 &   1.61  &   2.25 &  1.77 \\
\hline\noalign{\smallskip}
Group 2:     &      &         &        &      \\
HS\,Aur      & 1.61 &   1.62  &   1.69 &  1.63 \\
RW\,Lac      & 1.58 &   1.60  &   1.76 &  1.60 \\
$\alpha$\,Cen& 1.58 &   1.61  &   2.06 &  1.78 \\ 
V568\,Lyr    & 1.59 &   1.64  &   2.02 &  1.42 \\
\hline\noalign{\smallskip}
Exception:   &      &         &        &      \\
VZ\,Cep      & 1.39 &   1.59  &   2.34 &  2.07 \\
\noalign{\smallskip}
\hline
\end{tabular}
\end{minipage}
\end{table}

%% file: v636_rv.tex
\begin{table*}
\caption[]{\label{tab:v636_rv}
Radial velocities of \V6\ and residuals from the final spectroscopic orbit
(in units of \kms).
$\sigma$ is the internal velocity error.
}
\begin{center}
\begin{tabular}{llrrrrrrrr}
\hline\noalign{\smallskip}
HJD       & $Phase$ & $RV_p$  &$\sigma_p$ & $O-C$ & HJD       &  Phase  & $RV_s$  & $\sigma_s$ &$O-C$ \\
$-$ 2\,400\,000&    &         &         &       & $-$ 2\,400\,000&    &         &         &    \\
\hline\noalign{\smallskip}
46515.783 &  0.4278 &  -59.46 &    0.64 &  1.15 & 46515.786 &  0.4285 &   41.69 &    1.68 & -0.74 \\
46516.730 &  0.6489 &   52.18 &    0.59 &  0.12 & 46516.727 &  0.6482 &  -95.99 &    1.13 & -0.30 \\
46517.736 &  0.8837 &   27.81 &    0.59 & -0.65 & 46517.741 &  0.8849 &  -65.81 &    1.67 &  0.59 \\
46518.796 &  0.1312 &  -56.07 &    0.49 & -0.05 & 46518.800 &  0.1321 &   37.31 &    1.57 & -0.17 \\
46520.719 &  0.5800 &   24.24 &    0.65 &  0.24 & 46520.725 &  0.5815 &  -60.52 &    1.01 &  1.76 \\
47219.878 &  0.7846 &   54.94 &    0.46 &  0.14 & 47219.888 &  0.7870 &  -96.58 &    0.82 &  2.18 \\
47221.904 &  0.2576 &  -83.06 &    0.41 &  0.12 & 47221.897 &  0.2559 &   70.14 &    1.08 & -0.20 \\
47222.899 &  0.4898 &  -29.04 &    0.50 &  0.01 & 47222.899 &  0.4898 &    3.04 &    0.77 & -0.89 \\
47223.870 &  0.7165 &   61.08 &    0.40 & -0.21 & 47223.881 &  0.7190 & -107.41 &    0.94 & -0.11 \\
          &         &         &         &       & 47223.894 &  0.7221 & -107.61 &    0.83 & -0.33 \\
47545.853 &  0.8769 &   30.61 &    0.50 & -0.04 & 47545.861 &  0.8788 &  -70.20 &    0.71 & -1.38 \\
47546.872 &  0.1148 &  -50.93 &    0.43 &  0.17 & 47546.878 &  0.1162 &   30.76 &    0.97 & -0.85 \\
47547.853 &  0.3438 &  -83.84 &    0.44 & -0.55 & 47547.859 &  0.3452 &   69.18 &    1.01 & -1.31 \\
47549.889 &  0.8190 &   46.96 &    0.41 & -0.33 & 47549.894 &  0.8202 &  -89.86 &    0.83 & -0.17 \\
47602.861 &  0.1843 &  -70.53 &    0.44 & -0.35 & 47602.868 &  0.1859 &   54.29 &    0.81 & -0.74 \\
47603.837 &  0.4121 &  -66.89 &    0.39 & -0.13 & 47603.829 &  0.4103 &   51.57 &    0.74 &  0.39 \\
47605.826 &  0.8764 &   31.10 &    0.44 &  0.29 & 47605.820 &  0.8750 &  -70.84 &    0.80 & -0.53 \\
47607.826 &  0.3433 &  -83.04 &    0.39 &  0.31 & 47607.819 &  0.3416 &   71.49 &    0.90 &  0.47 \\
47608.810 &  0.5730 &   20.32 &    0.42 &  0.11 & 47608.816 &  0.5744 &  -57.87 &    0.77 & -0.23 \\
47609.836 &  0.8125 &   49.29 &    0.44 &  0.40 & 47609.840 &  0.8134 &  -91.01 &    0.93 &  0.73 \\
47662.741 &  0.1621 &  -64.69 &    0.41 & -0.05 & 47662.748 &  0.1637 &   49.92 &    0.85 &  1.65 \\
47663.671 &  0.3792 &  -77.34 &    0.42 & -0.55 & 47663.667 &  0.3782 &   62.24 &    0.78 & -0.74 \\
47663.781 &  0.4048 &  -69.39 &    0.43 & -0.08 & 47663.786 &  0.4060 &   54.02 &    0.85 &  1.01 \\
47664.766 &  0.6348 &   47.90 &    0.45 &  0.11 & 47664.771 &  0.6359 &  -91.61 &    0.79 & -0.47 \\
47665.647 &  0.8404 &   41.99 &    0.43 &  0.37 & 47665.652 &  0.8416 &  -85.02 &    0.84 & -2.35 \\
47665.774 &  0.8701 &   32.49 &    0.44 & -0.32 & 47665.779 &  0.8712 &  -74.63 &    1.37 & -2.86 \\
47669.651 &  0.7751 &   56.50 &    0.50 &  0.04 & 47669.658 &  0.7767 &  -99.40 &    0.87 &  1.60 \\
47669.735 &  0.7947 &   52.81 &    0.43 & -0.03 & 47669.741 &  0.7961 &  -95.27 &    0.76 &  1.26 \\
47671.639 &  0.2391 &  -80.55 &    0.42 &  0.25 & 47671.635 &  0.2382 &   66.11 &    0.89 & -1.37 \\
47671.762 &  0.2678 &  -84.03 &    0.41 &  0.16 & 47671.768 &  0.2692 &   72.54 &    0.94 &  0.57 \\
47671.827 &  0.2830 &  -84.91 &    0.45 &  0.34 & 47671.833 &  0.2844 &   70.99 &    1.01 & -2.21 \\
47673.621 &  0.7018 &   60.76 &    0.44 & -0.03 & 47673.628 &  0.7034 & -107.20 &    0.98 & -0.42 \\
47673.684 &  0.7165 &   61.45 &    0.42 &  0.16 & 47673.691 &  0.7181 & -108.41 &    0.88 & -1.11 \\
47673.761 &  0.7345 &   61.14 &    0.41 &  0.24 & 47673.770 &  0.7366 & -106.04 &    0.88 &  0.62 \\
47673.793 &  0.7419 &   60.37 &    0.43 & -0.06 & 47673.802 &  0.7440 & -106.00 &    0.91 &  0.02 \\
47675.681 &  0.1827 &  -69.37 &    0.42 &  0.42 & 47675.687 &  0.1841 &   55.57 &    0.72 &  1.08 \\
47676.773 &  0.4376 &  -56.04 &    0.43 &  0.35 & 47676.781 &  0.4394 &   37.11 &    0.85 &  0.56 \\
47699.521 &  0.7476 &   59.71 &    0.49 & -0.25 & 47699.526 &  0.7488 & -108.24 &    1.19 & -2.72 \\
47701.670 &  0.2493 &  -82.79 &    0.56 & -0.60 & 47701.675 &  0.2504 &   68.79 &    1.81 & -0.75 \\
\noalign{\smallskip}
\hline
\end{tabular}
\end{center}
\end{table*}

%% file: clausen.bbl
\begin{thebibliography}{}
\bibitem[1996]{alonso96}                               
Alonso, A., Arribas, S., \& Mart\'{\i}nez-Roger, C.
1996, \aap, 313, 873
\bibitem[1998]{baraffe98}                               
Baraffe, I., Chabrier, G., Allard, F., \& Hauschildt, P.
1998, \aap, 337, 403
\bibitem[1979]{b79}                                    
Baranne, A., Mayor, M., \& Poncet, J.-L. 1979,
Vistas in Astron., 23, 279
\bibitem[2004]{bruntt04}                               
Bruntt, H., Bikmaev, I.~F., Catala, C. et al. 2004,
\aap, 425, 683
\bibitem[2008]{bruntt08}                               
Bruntt, H., De Cat, P., \& Aerts, C. 2008,
\aap, 478, 487
\bibitem[2003]{cb03}                                    
Carrier, F. \& Bouran, G.. 2003, \aap, 406, L23
\bibitem[2007]{cgb07}
Chabrier, G., Gallardo, J., \& Baraffe, I. 2007,
\aap, 472, L17
\bibitem[1995]{claret95}
Claret, A. 1995, \aaps, 109, 441
\bibitem[1997]{claret97}
Claret, A. 1997, \aaps, 125, 439
\bibitem[2000]{c00}                                     
Claret, A., 2000, A\& A,363, 1081
\bibitem[2004]{c04}                                 
Claret, A., 2004, A\& A,424, 919
\bibitem[2005]{c05}                                 
Claret, A., 2005, A\& A,440, 647 
\bibitem[2006]{c06}                                 
Claret, A., 2006, A\& A,453, 769 
\bibitem[2007]{c07}                                 
Claret, A., 2007, A\& A,467, 1389
\bibitem[1997]{ac97}                                     
Claret, A. \& Cunha, N.~C.~S. 1997,
\aap, 318, 187
\bibitem[1995]{ac95}                                     
Claret, A., Gim\'enez, A., \& Cunha, N.~C.~S.
1995, \aap, 299, 724
\bibitem[1999]{granada99}
Clausen,  J.~V., Baraffe,  I., Claret,  A., \&
VandenBerg,  D.~B. 1999b,
in Theory and Tests of Convection in Stellar
Structure, ed. A. Gim\'enez, E. F. Guinan, \& B. Montesinos,
ASP Conference Series, 173, 265
\bibitem[2001]{jvcetal01}                              
Clausen, J.~V., Helt, B.~E., \& Olsen, E.~H. 2001,
\aap, 374, 980
\bibitem[2008a]{jvcetal08}                              
Clausen, J.~V., Vaz, L.~P.~R., Garc\'ia, J.~M., et al.
2008a, \aap, 487, 1081 (CVG08) 
\bibitem[2008b]{avw08}                                 
Clausen, J.~V., Torres, G., Bruntt, H., et al. 2008b,
\aap, 487, 1095 (CTB08)
\bibitem[1981]{csh81}
Cox, A.~N., Shaviv, G., \& Hodson, S.~W. 1981, \apj, 245, L37
\bibitem[2000]{da00}                                  
D'Antona, F., Ventura, P., \& Mazzitelli, I. 2000,
\apj, 543, L77
\bibitem[2004]{ddr04}
Dawson, P.~C., \& De Robertis, M.~M. 2004, \aj, 127, 2909
\bibitem[2004]{yale04}
Demarque, P., Woo, J.-H., Kim, Y.-C., \& Yi, S.~K.   
2004, \apjs, 155, 667
\bibitem[2004]{eggenberger04}                           
Eggenberger, P., Charbonnel, C., Talon, S. et al. 2004,
\aap, 417, 235
\bibitem[1997]{hip97}
ESA 1997, The Hipparcos and Tycho Catalogues, 
ESA SP-1200
\bibitem[1993]{be93}                                   
Edvardsson, B., Andersen, J., Gustafsson, B. et al.
1993, A\&A, 275, 101
\bibitem[2004]{sbop}                                   
Etzel, P.~B. 2004, SBOP: 
Spectroscopic Binary Orbit Program
San Diego State University)
\bibitem[1996]{flower96}                               
Flower, P.~J. 1996, \apj, 469, 355
\bibitem[1969]{g69}
Gabriel, M. 1969, in Low-Luminosity Stars,
ed. S.S. Kumar, New York: Gordon \& Breach, 267
\bibitem[1985]{agc85}                                  
Gim\'enez A. 1985, \apj, 297, 405
\bibitem[1991]{agc91}                                  
Gim\'enez, A., Reglero, V., de Castro, E., \&
Fern\'andez-Figueroa, M.~J. 1991, \aap, 248, 563
\bibitem[1998]{gs98}                                   
Grevesse, N, \& Sauval, A.~J. 1998,
\ssr, 85, 161
\bibitem[1996]{gns96}                                  
Grevesse, N., Noels, A., \& Sauval, A.~J. 1996,
in Cosmic Abundances, eds. S.S. Holt \& G. Sonneborn
(San Francisco: ASP), 117
\bibitem[2007]{gas07}                                   
Grevesse, N., Asplund, M., \& Sauval, A.~J. 2007,
\ssr, 130, 105
\bibitem[2008]{gchf08}
Grundahl, F., Clausen, J.~V., Hardis, S., \& Frandsen, S.  
2008, \aap, 492, 171
\bibitem[2008]{marcs08}                               
Gustafsson, B., Edvardsson, B., Eriksson, K. et al.
2008, \aap, 486, 951
\bibitem[1994]{hall94}                                 
Hall, D.~S. 1994, \memsai, 65, 73
\bibitem[2002]{heiter02}                               
Heiter, U., Kupka, F., van't Menneret, C. et al.
2002, A\&A, 392, 619
\bibitem[1996]{henry96}                                
Henry, T.~J., Soderblom, D.~R., Donahue, R.~A., \&
Baliunas, S.~L. 1996, \aj, 111, 439
\bibitem[1958]{h58}
Hoffmeister, C. 1958,                                  
Ver\"offentlichungen der Sternwarte in Sonneberg, 3, 439
\bibitem[2007]{holmberg07}                             
Holmberg, J., Nordstr\"om, B., \& Andersen, J. 2007
\aap, 475, 519
\bibitem[1978]{houk78}                                 
Houk, N. 1978, Michigan Catalogue of two-dimensional 
spectraltypes for HD stars. Vol. 2, Dep. Astron., Univ. 
Michigan, Ann Arbor, Michigan, USA
\bibitem[1973]{h73}
Hoxie, D.~T. 1973, \aap, 26, 437
\bibitem[1981]{hut81}                                  
Hut, P. 1981, \aap, 99, 126
\bibitem[2005]{hk05}                                   
Kjeldsen, H., Bedding, T.~R., Butler, R.~P., et al.
2005, \apj, 635, 1281
\bibitem[1999]{kupka99}                                
Kupka, F., Piskunov, N., Ryabchikova, T.~A.,
Stempels, H.~C., \& Weiss, W. 1999, A\&AS, 138, 119
\bibitem[2005]{ltcv05}
Lacy, C.~H.~S., Torres, G., Claret, A., \& Vaz, L.~P.~V. 
2005, \aj, 130, 2838
\bibitem[1998]{al98}                                   
Larsen, A. 1998,
Master Thesis, Copenhagen University
\bibitem[2007]{lm07}
L\'opez-Morales, M. 2007, \apj, 660, 732 
\bibitem[2005]{lmir05}
L\'opez-Morales, M., \& Ribas, I. 2005, \apj, 631, 1120
\bibitem[1999]{hgl99}                                     
Ludwig, H:-G., Freytag, B, \& Steffen, M. 1999,
\aap, 346, 111
\bibitem[1985]{m85}                                    
Mayor, M. 1985, in Stellar Radial Velocities,
ed. A.~G.~D. Philip \& D.~W. Latham
(Schenectady, New York: L. Davis Press),
IAU Coll. 88, 35
\bibitem[2008]{mrj08}
Morales, J.~C., Ribas, I., \& Jordi, C. 2008,
\aap, 478, 507
\bibitem[1997]{motch97}                                 
Motch, C., Guillout, P., Haberl, F., et al. 1997,
\aaps, 122, 201 
\bibitem[2004]{gencph04}
Nordstr\"om, B., Mayor, M., Andersen, J., et al.
2004, \aap, 418, 989
\bibitem[1988]{eho88}                                  
Olsen, E.~H. 1988, A\&A, 189, 173
\bibitem[2004]{basti04}                                
Pietrinferni, A., Cassisi, S., \& Salaris, M.
2004, Mem. Soc. Astron. Italiana, 75, 170
\bibitem[1966]{dmp66}                                  
Popper, D.~M. 1966, AJ, 71, 175
\bibitem[1997]{dmp97}                         
Popper, D.~M. 1997, AJ, 114, 1195
\bibitem[1998]{dmp98}                                
Popper, D.~M. 1998, PASP, 110, 919 
\bibitem[1986]{dmp86}                                  
Popper, D.~M., Lacy, C.~H., Frueh, M.~I., \&
Turner, A.~E. 1986, \aj, 91, 383
\bibitem[2005]{rm05}                                   
Ram\'\i rez, I., \& Mel\'endez, J. 2005, AJ, 626, 465
\bibitem[2003]{ir03}
Ribas, I. 2003, \aap, 398, 239
\bibitem[2008]{rmjbcg08}
Ribas, I., Morales, J.~C., Jordi, C. et al. 2008,
Mem. Soc. Astron. Italiana, 79, 562
\bibitem[2004]{stassun04}                               
Stassun, K.~G., Mathieu, R.~D., Vaz, L.~P.~R. et al.
2004, \apjs, 151, 357
\bibitem[2004]{sms04}                                 
Siviero, A., Munari, U., Sordo, R. et al. 2004,
\aap, 417, 1083
\bibitem[1997]{tt97}                                  
Tassoul, M. \& Tassoul, J.-L. 1997, \apj, 481, 363
\bibitem[2002]{tr02}
Torres, G., \& Ribas, I. 2002, \apj, 567, 1140
\bibitem[2009]{tl09}                                   
Torres, G., \& Lacy, C.~H.~S. 2009,
\aj, 137, 507
\bibitem[2006]{wt06}                                   
Torres, G., Lacy, C.H., Marschall, L.~A.,
Sheets, ,H.~A, \& Mader, J.~A. 2006, \aj, 640, 1018
\bibitem[1996]{vp96}                                   
Valenti, J., \& Piskunov, N. 1996, A\&AS, 118, 595
\bibitem[2006]{vr06}                                    
VandenBerg, D.~A., Bergbusch, P.~A., \& Dowler, P.~D. 2006
\apjs, 162, 375
\bibitem[1993]{vh93}                                    
Van Hamme, W., 1993, AJ, 106, 2096
\bibitem[2007]{vl07}                                  
van Leeuwen, F. 2007, Hipparcos, the new reduction
of the raw data (Dordrecht: Springer)
\bibitem[1999]{rosat99}                               
Voges, W., Aschenbach, B., Boller, T. et al. 1999
\bibitem[2000]{rosat00}                               
Voges, W., Aschenbach, B., Boller, T. et al. 2000
IAU Circ., 7432, 1
\bibitem[1977]{zahn77}                                  
Zahn, J.-P. 1977, \aap, 57, 383
\bibitem[1989]{zahn89}                                  
Zahn, J.-P. 1989, \aap, 220, 112
\bibitem[1998]{whl98}                                   
Wallace, L., Hinkle, K., \& Livingston, W. 1998,
An atlas of the spectrum of the solar photosphere 
from 13500 to 28000 \cm\ (3570 to 7405 A),
Tucon, AZ: NOAO
\bibitem[1967]{wolfe67}       
Wolfe, R.~H., Horak, H.~G., Storer, N.~W., 1967,    
in Modern Astrophysics: 
A Memorial to Otto Struve, ed.\ M.\ Hack, 251 
\bibitem[2007]{yildiz07}                        
Yildiz, M. 2007, \mnras, 374, 1264
\bibitem[2008]{yildiz08}                        
Yildiz, M. 2008, in The Art of Modelling Stars  
in the 21th Century, eds. L. Deng \& K.~L. Chan,
IAU Symp. 252, 183
\bibitem[2006]{yildiz06}                        
Yildiz, M., Yakut, K., Bakis, H., \& Noels, A. 
2006, \mnras, 368, 1941
\end{thebibliography}
